\documentclass[superscriptaddress,aps,prx,nofootinbib,notitlepage,10pt,longbibliography]{revtex4-2}
\usepackage{amsmath}
\usepackage{mathtools}
\usepackage{amssymb}
\usepackage{physics}
\usepackage{graphicx}
\usepackage{color}
\usepackage{listings}
\usepackage{float}
\usepackage{booktabs}
\usepackage{multirow}
\usepackage[table,xcdraw]{xcolor}
\usepackage{hyperref}
\hypersetup{
	pdfstartview={FitH},
	pdfnewwindow=true,
	colorlinks=true,
	linkcolor=blue,
	citecolor=blue,
	filecolor=blue,
	urlcolor=blue}
\usepackage{enumitem}
\usepackage{tikz}
\usetikzlibrary{quantikz2}

\definecolor{cUV}{HTML}{D9EAD3}  
\definecolor{cV}{HTML}{FCE5CD}   
\definecolor{cDir}{HTML}{C9DAF8} 

\usepackage{listings}

\newcommand{\nn}{\nonumber \\}

\newcommand{\MQ}{\affiliation{%
School of Mathematical and Physical Sciences,
Macquarie University, Sydney, NSW, Australia}}

\begin{document}
\author{Alicja Dutkiewicz}
\email{alicja.dutkiewicz@cwi.nl}
\affiliation{
             Google Quantum AI, 
             San Francisco, CA, USA}
\affiliation{
            QuSoft \& CWI, Amsterdam, the Netherlands
             }

\author{Alec F. White}
\email{alecwhite@google.com}
\affiliation{
             Google Quantum AI, 
             San Francisco, CA, USA}
             
\author{Guang Hao Low}
\affiliation{
             Google Quantum AI, 
             San Francisco, CA, USA}

\author{A. Eugene DePrince III}
\affiliation{
             Department of Chemistry and Biochemistry,
             Florida State University,
             Tallahassee, FL 32306-4390, USA}
\affiliation{
             Google Quantum AI, 
             San Francisco, CA, USA}
             
\author{Matthew P. Harrigan}
\affiliation{
             Google Quantum AI, 
             San Francisco, CA, USA}

\author{Marika Kieferova}
\affiliation{
             Google Quantum AI, 
             Sydney, New South Wales, AUS}
\affiliation{
             School of Computer Science, University of Technology Sydney, 
             Sydney, New South Wales, AUS}
             
\author{Ryan Babbush}
\affiliation{
             Google Quantum AI, 
             Los Angeles, CA, USA}

\author{Dominic W.~Berry}
\email{dominic.berry@mq.edu.au}
\MQ

\author{Nicholas C. Rubin}
\email{nickrubin@google.com}
\affiliation{
             Google Quantum AI, 
             San Francisco, CA, USA}

\title{Spectral amplification for ground-state energy estimation of electronic structure in first quantization}
\begin{abstract}
We demonstrate an asymptotic gate complexity improvement in first-quantized ground-state energy estimation of electronic structure Hamiltonians in a plane wave basis by employing the sum-of-squares spectral gap amplification protocol.  The improvement relies on identifying a sum-of-squares representation of the Hamiltonian which provides a lower bound certificate and low cost block encoding that leads to a provably lower quantum phase estimation gate cost.  This is achieved by using a sum-of-squares operator generated by the total charge density operator resulting in a block encoding normalization improvement of $\lambda = \mathcal{O}\left(\eta\Delta^{-1.5}+\eta^{1.5}\Delta^{-1} \right)$ compared to prior work $\lambda = \mathcal{O}(\eta\Delta^{-2}+\eta^2\Delta^{-1})$ where $\eta$ is the number of electrons and $\Delta$ is the simulation grid spacing. The asymptotic reduction in block encoding normalization and similar block encoding costs to prior work is demonstrated to reduce resource estimates for materials and chemical systems by a factor of $2 - 44\times$ corresponding to the lowest cost estimates for \textit{ab initio} materials simulation.
\end{abstract}
\maketitle

\tableofcontents

\section{Introduction}

Simulating the electronic structure of chemical and material systems is a promising application target for quantum algorithms with the potential for significant impact in science and engineering. While most work on quantum algorithms for electronic structure has focused on algorithms that solve the ground-state problem in second quantization
~\cite{aspuru2005simulated,whitfield2011simulation,babbush2016exponentially,reiher2017elucidating,babbush2018encoding,berry2019qubitization,von2021quantum,lee2021even,goings2022reliably,rubin2023fault,ivanov2023quantum,oumarou2024accelerating,berry2025rapid,lowsossa2025}, a complementary approach targets a first-quantized representation that provides an exponential space advantage allowing simulations to target the large-basis-set limit
~\cite{babbush2019quantum,su2021fault,delgado2022simulating,zini2023quantum,rubin2024quantum,berry2024quantum,pocrnic2026efficient,da2025comprehensive,georges2025quantum}. In the first-quantized approach, each of the $\eta$ electrons is simulated using a separate qubit register, resulting in a total memory requirement that scales as $\mathcal{O}(\eta \log N)$, where $N$ is the number of basis functions. This is in contrast to the second-quantized representation, where qubits represent the occupancy of specific orbitals, and $\mathcal{O}(N)$ qubits are needed.

Despite the favorable memory scaling of first-quantized algorithms with respect to the number of basis functions, the gate counts for quantum algorithms formulated based on queries to a block encoding of the Hamiltonian end up being large in practice when using large plane-wave basis sets. This is true even if one uses physically motivated approximations to reduce the size of the basis \cite{zini2023quantum,berry2024quantum}. In this representation, an efficient implementation of block encoding relies on coherent arithmetic, incurring a modest $\text{polylog}$ Toffoli cost in $N$ and minimal ancilla overhead. The primary reason for the high gate costs in tasks such as ground-state energy estimation and time-evolution~\cite{eklund2026end,rubin2024quantum} is the block encoding normalization constant. Recent work in Ref.~\cite{georges2025quantum} attempts to minimize the block encoding normalization factor with a variety of space and time tradeoffs by using compact basis sets~\cite{georges2025quantum}. The goal of this work is to reduce the high cost of ground-state energy estimation in first quantization and maintain a favorable space advantage.

Reducing the cost of ground-state energy estimation in first quantization is particularly relevant to the quantum simulation of materials. Simulating materials on a quantum or classical computer is difficult because experimentally relevant results require converging to the thermodynamic limit. This means that in practice one must do calculations on increasingly large systems, which results in daunting costs for quantum algorithms. This is true in second quantization, where even efficient use of the translational symmetry of the crystal does not ameliorate the cost \cite{rubin2023fault,ivanov2023quantum}, and in first quantization where the gate counts can still be large despite favorable asymptotic scaling. This favorable scaling makes first-quantized algorithms a natural choice for materials, and recent work has improved on the basic algorithm by incorporating pseudopotentials of different kinds \cite{zini2023quantum,berry2024quantum,ivanov2025quantum}, a necessary step for practical applications. In this work, we aim to improve both the prefactor and the asymptotic scaling without introducing additional physical approximations. The implementation of pseudopotentials within our approach is left to future work.

Our tool will be the recently introduced Sum-of-Squares Spectral Amplification (SOSSA) framework \cite{lowsossa2025,king2026quantum} which has been shown to asymptotically reduce gate counts for ground-state energy estimation of molecular systems~\cite{lowsossa2025} and asymptotically improve gate costs for strongly-correlated model problems, such as the SYK model, at any temperature~\cite{king2026quantum}. Separately, similar gap amplification protocols have been recently applied to accelerate quantum Gibbs sampling~\cite{leng2026accelerating}. SOSSA is formulated in a block encoding query model and requires access to a square root of a non-negative representation of the Hamiltonian. The additional access provides information about the square root of the non-negative Hamiltonian~\cite{somma2013spectral, lowsossa2025} allowing one to obtain dynamics faster and reduce precision requirements in many estimation tasks.  
In the ground-state energy estimation setting using quantum phase estimation (QPE), the non-negative form of the Hamiltonian is obtained by determining an equivalent sum of squares representation up to a constant shift, 
\begin{align}\label{eq:sos_h_sossa}
H+\beta\mathbb{I} = \sum_{\alpha}O_{\alpha}^{\dagger}O_{\alpha} = H_{\text{SOSSA}}^\dagger H_{\text{SOSSA}},
\end{align}
where $-\beta$ is a lower bound on the ground-state energy and $O_{\alpha}$ are operators.
Then a block encoding of $H_{\text{SOSSA}}/\sqrt{\lambda_{\text{SOS}}}$ is used to construct a walk operator with eigenphases
\begin{align}\label{eq:sossa_phases}
\phi_k = \cos^{-1}\left(2\frac{E_k+\beta}{\lambda_{\mathrm{SOS}}}-1\right) ,
\end{align}
where $E_{k}$ are the eigenvalues of $H$. In this framework, assuming we are interested in ground-state energy estimation, the cost of phase estimation scales with $\lambda_{\text{eff}}$, an effective block encoding normalization,
\begin{equation}
\label{eq:lambda_eff}
    \lambda_{\mathrm{eff}} \equiv
    \left|\frac{\dd \phi_0}{\dd E_0}\right
    |^{-1} = \sqrt{(\lambda_{\mathrm{SOS}}-E_{\text{gap}})E_{\text{gap}}} \, ,
\end{equation}
where $E_{\text{gap}}=E_0+\beta$ can be arbitrarily small as the bound $-\beta$ becomes tighter. To reduce the total cost of the algorithm we must account for the cost of implementing a block encoding of $H_{\mathrm{SOSSA}}$ which can increase rapidly as the lower bound certificate $\beta$ improves. Fewer block encoding calls but higher block encoding cost is the core tradeoff that needs to be accounted for when demonstrating SOSSA can provide reduced end-to-end costs.
State-of-the-art end-to-end algorithm costs for ground-state energy estimation of a second-quantized electronic structure Hamiltonian \cite{lowsossa2025, low2026denser} leveraged an efficient semidefinite programming preprocessing step to produce the near-frustration-free representation~\cite{rubin2026near}. Extending the same non-commutative polynomial optimization procedure to a first-quantized Hamiltonian presents a numerical challenge; large basis sets lead to very large semidefinite programs out of reach of numerical solvers. Furthermore, such a solution would likely incur a large gate cost spoiling the exponential space advantage of first quantization. 

In this work, we resolve these difficulties by constructing an analytical SOS representation of the first-quantized electronic structure Hamiltonian in a plane-wave basis that is implemented with similar cost to prior methods, and provides a sufficiently tight lower-bound certificate to prove an asymptotic advantage when applying phase estimation to a low-energy state. The key component of the SOS representation is constructed by noting that a low-frustration representation can be generated by the total charge density operator. Squaring produces the original Hamiltonian and a self-interaction term which is converted to an identity term by an additional SOS generator. The resulting block encoding normalization has the same scaling as the original linear combination of unitaries (LCU) in Ref.~\cite{su2021fault}. By compilation of the block encoding, we are able to verify that the asymptotic advantage obtained by gap amplification reduces total Toffoli complexity by $2-44\times$ compared to prior methods of Ref.~\cite{su2021fault} and preserves the space advantage. The full compilation demonstrates that many of the coherent arithmetic subroutines remain the same between the two implementations and the full walk operator cost is similar to the LCU method.  This isolates the observed reduced end-to-end Toffoli complexity to the improvement in the effective number of queries to the walk operator due to gap amplification.

The remainder of the paper is organized as follows.
In Sec.~\ref{sec:fq_sos} we describe the SOS form of the first-quantized Hamiltonian in detail.
In Sec.~\ref{sec:asymptotic_advantage} we quantify the resulting asymptotic speedup, showing that it is $\lambda_{\mathrm{eff}} = \mathcal{O}\left(\eta\Delta^{-1.5}+\eta^{1.5}\Delta^{-1} \right)$.
In Sec.~\ref{sec:logical_resources} we provide detailed logical resource estimates, which we use to numerically confirm the asymptotic speedup. We show that this translates to a speedup of $2 - 44\times$ over the previous state-of-the-art on a number of benchmark systems.
In Sec.~\ref{sec:remarks} we conclude with a summary of the main results and some comments on future directions.

\section{First-Quantized Hamiltonian and a Sum-of-Squares Representation}\label{sec:fq_sos}

\subsection{Plane-wave Hamiltonian}
The non-relativistic Hamiltonian for a molecular or material system has the form
\begin{equation}
\label{eq:hamiltonian_terms}
H = T+V+U,
\end{equation}
where the kinetic energy ($T$), electron-electron interaction ($V$), and electron-nuclear interaction ($U$), can be resolved in a plane-wave basis as
\begin{align}
    T &= \sum_{j=1}^\eta \sum_{p\in G} \frac{\|k_p\|^2}{2} \ket{p}\bra{p}_j, \label{eqn:T}\\
    V &= \frac{2\pi}{\Omega} \sum_{i\ne j=1}^\eta \sum_{p,q\in G} \sum_{\substack{\nu\in G_0\\q-\nu\in G\\p+\nu\in G}} \frac 1{\|k_\nu\|^2} \ket{p+\nu}\bra{p}_i \ket{q-\nu} \bra{q}_j , \label{eqn:V}\\
    U &= -\frac{4\pi}{\Omega}\sum_{\ell=1}^{L} \sum_{j=1}^{\eta}\sum_{p \in G} \sum_{\substack{\nu \in G_0\\ p-\nu\in G}}\left(\zeta_{\ell}\frac{e^{ik_{\nu}\cdot R_{\ell}}}{\|k_{\nu}\|^{2}} \right) \ket{p-\nu}\bra{p}_j .\label{eqn:U}
\end{align}
Note that we are making the Born-Oppenheimer approximation and using the notation of Ref.~\cite{su2021fault}. In particular, $\eta$ is the number of electrons, $L$ is the number of nuclei, $\Omega$ is the volume of the real-space unit cell, and $\{R_{\ell}\}$ are the position vectors of the nuclei in the computational unit cell with charges $\{ \zeta_{\ell}\}$. The sum of the nuclear charges is the total nuclear charge, $Z_{\mathrm{tot}}$, which is equal to the number of electrons ($\eta$) for a charge-neutral system. The plane-wave basis is also the natural representation for the uniform electron gas (UEG), where the Hamiltonian is identical except that there is no electron-nuclear interaction:
\begin{equation}
    H_{\mathrm{UEG}} = T + V.
\end{equation}
In this work, we assume a cubic unit cell which defines a cubic reciprocal lattice given by
\begin{equation}\label{eqn:k_nu_def}
    k_{p} = \frac{2\pi p}{\Omega^{1/3}}, \qquad p \in G, \qquad G \equiv \left[ -\frac{N^{1/3} - 1}{2}, \frac{N^{1/3} - 1}{2}\right]^3.
\end{equation}
Here $N$ is the total number of plane-wave basis functions and $N^{1/3}$ is an odd integer for a cubic lattice. We also use $G_0$ to indicate the set of all non-zero differences in $G$,
\begin{equation}
    G_0 \equiv [-(N^{1/3} - 1), N^{1/3} - 1]^3 \backslash  \{(0, 0, 0)\}.
\end{equation}
The product structure of the computational basis means that it does not correspond to a precise momentum cutoff. However, to compare to classical calculations we can define a momentum cutoff, $k_{\mathrm{cutoff}}$, such that 
\begin{equation}
    \left(\frac{2\pi}{\Omega^{1/3}}\frac{N^{1/3} - 1}{2}\right) < k_{\mathrm{cutoff}}.
\end{equation}
In other words, $N^{1/3}$ is chosen so that the maximum absolute momentum of any 1-dimensional plane wave is less than $k_{\mathrm{cutoff}}$ and the product basis used in the quantum calculation is a superset of the basis formed from a spherical $k_{\mathrm{cutoff}}$. The real-space grid has the same number of grid points, which defines an effective grid spacing, $\Delta$, with units of length such that
\begin{equation}\label{eqn:grid_spacing}
    \Delta^{-1} \equiv \left(\frac{N}{\Omega}\right)^{1/3}.
\end{equation}

\subsection{The LCU representation}
In Ref.~\cite{su2021fault} it was shown that this Hamiltonian (Eqs.~\eqref{eqn:T}-\eqref{eqn:U}) can be rewritten as a LCU:
\begin{align}
    T &= \sum_{j=1}^\eta \sum_{w \in \{x,y,z\}} \sum_{r,s =0}^{n_p-2} \sum_{b\in \{0,1\} } \frac{\pi^2}{\Omega^{2/3}}2^{r+s} \left(\sum_{p\in G} (-1)^{b(1\oplus p_{w,s}p_{w,r})} \ket{p}\bra{p}_j \right),\\
    V &= \sum_{\nu \in G_0} \sum_{i\neq j} \sum_{b\in \{ 0,1 \} } \frac{1}{4\pi\Omega^{1/3}}\frac{1}{\rVert\nu \lVert^2}\left( \sum_{p,q\in G}(-1)^{b(1\oplus(p+v\in G \land q-\nu \in G))} \ket{p+\nu}\bra{p}_i \ket{q-\nu}\bra{q}_j \right), \\
    U &=\sum_{\nu \in G_{0}}\sum_{\ell=1}^{L}\frac{\zeta_{\ell}}{2\pi\Omega^{1/3} \|\nu\|^{2}}
    \sum_{j=1}^{\eta}\sum_{b\in\{0, 1\}}\left( -e^{i k_{\nu} \cdot R_{\ell}} \sum_{p \in G}  \left(-1\right)^{b[(p - \nu) \notin G]} \vert p - \nu\rangle\langle p\vert_{j} \right).
\end{align}
Here we have used
\begin{equation}
\label{eq:n-qubits-per-dim}
    n_p \equiv \left\lceil \log_2\left(N^{1/3} + 1\right) \right\rceil,
\end{equation}
and $\{p_{w,s}$,$p_{w,r}\}$ are bits in the representation of the momentum vector in sign-magnitude form as described in Ref.~\cite{su2021fault} and in Appendix~\ref{app:block_encoding}.
The LCU normalization factors for these terms were found to be (see Eq.~(25) of Ref.~\cite{su2021fault})
\begin{align}
    \label{eq:lambda-standard}
    \lambda_{\mathrm{LCU}} &= \lambda_T + \lambda_V + \lambda_U,\\
    \lambda_T & = \frac{3\pi^2}{2} \frac{\eta (N^{1/3}-1)^2}{\Omega^{2/3}} \lesssim \frac{3\pi^2}{2} \frac{\eta N^{2/3}}{\Omega^{2/3}},\label{eqn:lambda_T} \\
    \lambda_V &= \frac{\eta(\eta-1)}{2\pi\Omega^{1/3}}\lambda_{\nu}
    \lesssim 8 \frac{\eta^2 N^{1/3}}{\pi\Omega^{1/3}},
    \label{eq:lambdaV} \\
    \lambda_{U} &= \frac{\eta\sum_{\ell}\zeta_{\ell}}{\pi \Omega^{1/3}} \lambda_{\nu}, \label{eq:lambdaU}\\
    \lambda_{\nu} &= \sum_{\nu \in G_{0}} \frac{1}{\|\nu\|^{2}} ,
\end{align}
with asymptotic scaling
\begin{equation}
 \lambda_{\mathrm{LCU}} = \mathcal{O}(\eta N^{2/3} \Omega^{-2/3} + \eta^2 N^{1/3}\Omega^{-1/3}) = \mathcal{O}(\eta\Delta^{-2}+\eta^2\Delta^{-1}).
\end{equation}
Note that for the UEG, $\lambda_{\mathrm{LCU}} = \lambda_T + \lambda_V$, so the asymptotic scaling is the same. It is shown in Ref.~\cite{babbush2019quantum,berry2024quantum} that
\begin{equation}
    \lambda_{\nu} \sim 24 N^{1/3} \left[\mathrm{Ti}_2 \left(3-\sqrt 8\right) - C + \frac{\pi} 2 \log\left( 1+\sqrt 2 \right) \right] \approx 15.3482\times  N^{1/3},
\end{equation}
where Ti$_2$ is the Lewin inverse tangent integral and $C$ is the
Catalan constant.
It is also possible to evaluate the sum to higher order by expanding the sum in a series of integrals, to give
\begin{equation}
    \lambda_{\nu} = 24 N^{1/3} \left[\mathrm{Ti}_2 \left(3-\sqrt 8\right) - C + \frac{\pi} 2 \log\left( 1+\sqrt 2 \right) \right] -16.58775714002888\ldots
    +\frac{\sqrt 2 \arccot\left(\sqrt 2\right)}{N^{1/3}-1/2} + \mathcal{O}(1/N) \, .
\end{equation}
Assuming that $\Omega \propto \eta$ and using the definition of the grid spacing in Eq.~\eqref{eqn:grid_spacing},
\begin{equation}\label{eqn:lambda_lcu_scaling}
    \lambda_{\mathrm{LCU}} = \mathcal{O}(\eta^{1/3}N^{2/3}+\eta^{5/3}N^{1/3}) .
\end{equation}
The large value of $\lambda_{\mathrm{LCU}}$ (see for example Table 5 of Ref.~\cite{berry2024quantum}) is behind the high end-to-end Toffoli costs of simulating systems in first quantization.

\subsection{The sum-of-squares representation}
We can write the first quantized Hamiltonian plus a shift in a non-negative, sum-of-squares form as
\begin{equation}
\label{eq:sos_decomposition}
H + \beta \mathbb{I} = \sum_{j=1}^\eta \sum_{w\in \{x,y,z\} } T_{j,w}^\dagger T_{j,w} + \sum_{\nu\in G_0} C_{\nu}\tilde{V}_{\nu}^\dagger \tilde{V}_{\nu}+\sum_{j=1}^{\eta}\sum_{\nu\in G_0} C_{\nu}\tilde{X}_{j,\nu}^\dagger \tilde{X}_{j,\nu}
\end{equation}
where the positive shift, $\beta$, provides a lower bound of $-\beta$ on the energy of the ground state. 
The $T_{j,w}$ operators are chosen so that the first term analytically reproduces the kinetic energy operator and the $\tilde{V}_{\nu}$ operator is chosen to reproduce the electron-electron and electron-nuclear potential energy. The SOS form of $\tilde{V}_{\nu}^{\dagger}\tilde{V}_{\nu}$ produces a diagonal self-interaction term which is transformed into an identity (and thus a constant shift) by the SOS generated by $\tilde{X}_{j,\nu}$. 
The scaling constants $C_\nu$ are defined to simplify the notation. Specifically, these operators have the following form:
\begin{align}
    T_{j,w} &\equiv \frac {\sqrt{2}\pi}{\Omega^{1/3}}\sum_{p\in G} |p_w|
     \ket{p}\bra{p}_j, \label{eq:t_sos_gen}\\
     \tilde{V}_{\nu} &\equiv  \left(\sum_{j=1}^\eta \sum_{\substack{p\in G: \\p-\nu \in G}}  \ket{p-\nu} \bra{p}_j\right) -\sum_{\ell=1}^{L}\zeta_{\ell}e^{-ik_{\nu}\cdot R_{\ell}}\mathbb{I} \, , \label{eq:v_sos_gen}\\
    \tilde{X}_{j,\nu} &\equiv 
    \sum_{\substack{p \in G: \\p-\nu \notin G}}  \ket{p - \nu} \bra{p}_j, \label{eq:x_sos_gen}\\
    C_{\nu} &\equiv \frac{2\pi}{\Omega \|k_{\nu}\|^{2}} =
    \frac{1}{2\pi\Omega^{1/3} \|\nu\|^{2}}.
\end{align}
Note that the effect of applying $\tilde{X}_{j,\nu}^\dagger \tilde{X}_{j,\nu}$ is that $p$ is unchanged, and there is just a check that $p-\nu \notin G$.
We could alternatively have used
$\tilde{X}_{j,\nu} = \sum_{\substack{p \in G: \\p-\nu \notin G}}  \ket{p} \bra{p}_j$,
but our choice here enables a more efficient block encoding.
This is because checking $p-\nu \notin G$ is more conveniently performed after the subtraction. In addition, we also note that
there is a negative sign in $e^{-ik_{\nu}\cdot R_{\ell}}$, because the sign is flipped in the Hermitian conjugate in $\tilde{V}_{\nu}^\dagger \tilde{V}_{\nu}$.
The constant shift arises entirely from the interaction terms, because the kinetic energy operators $T_{j,w}$ reproduce $T$ with no shift.
It can be decomposed into self-interaction and potential contributions as
\begin{align}
    \label{eq:beta-def}
    \beta&=\eta \lambda_{\text{out}} + \beta_{\text{nuc}},\\
    \label{eq:lambda-out}
    \lambda_{\text{out}} &\equiv \sum_{\nu\in G_{0}}C_{\nu},\\
    \label{eq:beta-nuc}
    \beta_{\text{nuc}} &\equiv \sum_{\nu \in G_0}C_\nu
     \left|\sum_{\ell}\zeta_{\ell} e^{ik_{\nu}\cdot R_{\ell}} \right|^2.
\end{align}

The rectangular block encoding of each term leads to normalization factors (not accounting for minor success probability adjustments due to compilation choices) of
\begin{align}
    \lambda_{T_{j,w}} &= \sum_{r =0}^{n_p-2} \frac{\sqrt{2}\pi}{\Omega^{1/3}} 2^{r} = \frac{\sqrt{2}\pi}{\Omega^{1/3}} \sum_{r =0}^{n_p-2} 2^{r} = \sqrt{\frac{2\pi^2}{\Omega^{2/3}}}(2^{n_p-1}-1)
    ,\\
    \lambda_{\tilde{V}_\nu} &= \eta + Z_{\text{tot}},\\
    \lambda_{\tilde{X}_{j,\nu}} &= 1.
\end{align}
In the block encoding, each square is implemented by a single step of oblivious amplitude amplification, which gives a block encoding of 
\begin{equation}
    2\frac{H + \beta \mathbb{I}}{\lambda_{\text{SOS}}} - \mathbb{I} .
\end{equation}
Here $\lambda_{\text{SOS}}$ is defined as
\begin{align}
    \lambda_{\mathrm{SOS}} &= \sum_{j=1}^\eta \sum_{w\in \{x,y,z\} }\lambda_{T_{j,w}}^2 + \sum_{\nu\in G_0} C_{\nu}\lambda_{\tilde{V}_{\nu}}^2+\sum_{j=1}^{\eta}\sum_{\nu\in G_0} C_{\nu}\lambda_{\tilde{X}_{j,\nu}}^2 \nn
    &= \frac{6 \pi^2\eta}{\Omega^{2/3}} (2^{n_{p} - 1} - 1)^2
    + \left(\eta + Z_{\text{tot}}\right)^{2}\sum_{\nu \in G_{0}}C_{\nu} + \eta\sum_{\nu\in G_{0}}C_{\nu} \nn
    &= \lambda_{T} + \left(\left(\eta + Z_{\text{tot}}\right)^{2} + \eta\right)\lambda_{\text{out}}, \label{eqn:lambda_SOS}
\end{align}
where $\lambda_T$ is defined in Eq.~\eqref{eqn:lambda_T} and $\lambda_{\text{out}}$ is defined in Eq.~\eqref{eq:lambda-out}. 
Here we are choosing $N^{1/3}$ such that $N^{1/3}=2^{n_p}-1$ Eq.~\eqref{eq:n-qubits-per-dim}, so the expressions for $\lambda_T$ in terms of $N^{1/3}$ and $n_p$ are consistent.

The physical intuition behind the SOS generator we use here, $\tilde{V}_\nu$, is one that annihilates a state that has equivalent electronic and nuclear charge density everywhere in space.
To see this, the Fourier transform of the real-space electron density operator 
\begin{equation}
    \hat{\rho}_{\text{elec}}(r) = \sum_{j=1}^\eta \delta(r - \hat{r}_j),
\end{equation}
is
\begin{equation}
\hat{\rho}_{\text{elec}}(k_\nu) = \sum_{j=1}^\eta e^{-i k_\nu \cdot \hat{r}_j}
\end{equation}
when evaluated at a reciprocal lattice vector $k_\nu$. The operator $e^{-i k_\nu \cdot \hat{r}_j}$ acts as a translation operator, which has the form
\begin{equation}
e^{-i k_\nu \cdot \hat{r}_j} = \sum_p \ket{p-\nu}\bra{p}_j = U_{\text{shift},j}(\nu).
\end{equation}
The real-space static nuclear charge density 
\begin{equation}
\rho_{\text{nuc}}(r) = \sum_{\ell=1}^L \zeta_\ell \delta(r - R_\ell)
\end{equation}
has a Fourier transform 
\begin{align}
    \rho_{\text{nuc}}(k_\nu) &= \int_{\Omega} \left( \sum_{\ell=1}^L \zeta_\ell \delta(r - R_\ell) \right) e^{-i k_\nu \cdot r} dr \nonumber \\
    &= \sum_{\ell=1}^L \zeta_\ell e^{-i k_\nu \cdot R_\ell}. \label{eq:nuclear_fourier}
\end{align}
Thus, by combining the quantum electron density operator $\hat{\rho}_{\text{elec}}(k_\nu)$ with the classical nuclear density $\rho_{\text{nuc}}(k_\nu)$, the complete potential SOS generator evaluates the Fourier transform of the total charge density:
\begin{align}
    \tilde{V}_\nu \propto \hat{\rho}_{\text{elec}}(k_\nu) - \rho_{\text{nuc}}(k_\nu) = \hat{\rho}_{\text{total}}(k_\nu). \label{eq:total_charge_operator}
\end{align}
The cross term in the SOS gives the electron-nuclear interaction with the correct negative sign and there is a constant shift due to the nuclear self-interaction. For the UEG, there is no $\rho_{\mathrm{nuc}}$, but the uniform positive background is implicitly included by omitting the divergent $\nu = 0$ term in the sum, so the same intuition applies. The consequence of this SOS is that any state that is annihilated by $\tilde{V}_{\nu}$ satisfies $\matrixel{\psi}{\hat{\rho}_{\text{elec}}}{\psi} = \rho_{\text{nuc}}$.

Note that it is also possible to naturally generalise the sum of squares to the non-Born Oppenheimer case by simply replacing $\rho_{\text{nuc}}(k_\nu)$ with $\hat\rho_{\text{nuc}}(k_\nu)$.
In that case we would use
\begin{equation}
   \tilde{V}_{\nu}^{\text{(non-BO)}}= \left(\sum_{j=1}^\eta \sum_{\substack{p\in G: \\p-\nu \in G}}  \ket{p-\nu} \bra{p}_j\right) -\sum_{\ell=1}^{L}\zeta_{\ell}\sum_{\substack{p\in G: \\p-\nu \in G}}\ket{p-\nu} \bra{p}_{\eta+\ell} .
\end{equation}
There would also need to be $\tilde{X}_{\eta+\ell,\nu}$ included in the sum for the Hamiltonian, and $\beta$ would be modified to
\begin{equation}
    \beta^{\text{(non-BO)}} = \frac{1}{2\pi\Omega^{1/3}} \sum_{\nu \in G_0}\frac{1}{\lVert \nu \rVert^2}
    \left(\eta+ \sum_{\ell}\zeta_{\ell}^2 \right).
\end{equation}
Note that $\beta$ no longer has a quadratic dependence on the sum of $\zeta_\ell$ in this form, because that now appears in the Coulomb terms between nuclei.
The value of $\lambda_{\text{SOS}}$ is unchanged from the Born Oppenheimer case.

There are a variety of alternative SOS forms that one could derive to represent the Hamiltonian. As an example, we provide two SOS representations of the (Born Oppenheimer) Hamiltonian in Appendix~\ref{app:alt_sos_reps}.

\section{Asymptotic Advantage for SOSSA}\label{sec:asymptotic_advantage}
To explicitly analyze the asymptotic advantage of the SOSSA protocol over standard LCU methods, we evaluate the effective 1-norm $\lambda_{\text{eff}}$ (Eq.~\eqref{eq:lambda_eff}) that dictates the query complexity of the quantum walk operator. In this section, we study the scaling of $\lambda_{\text{eff}}$ with the number of electrons $\eta$ and the inverse grid spacing $\Delta^{-1}$ (Eq.~\eqref{eqn:grid_spacing}) under physically motivated assumptions.
Since $\lambda_{\text{eff}}$ is bounded by the geometric mean of the SOS norm and the gap:
\begin{equation}
\label{eq:lambda_eff_bound}
    \lambda_{\text{eff}} \leq \sqrt{\lambda_{\text{SOS}}(\beta + E_0)},
\end{equation}
we first analyze the asymptotic behavior of $\lambda_{\text{SOS}}$, $\beta$, and $E_0$.

The constituent 1-norms defined in Eq.~\eqref{eqn:lambda_T} and Eq.~\eqref{eq:lambda-out} have the following asymptotic behavior:
\begin{align}
    \lambda_T &= \mathcal{O}(\eta \Delta^{-2}) \label{eq:lambda_T_asymp} \\
    \lambda_{\text{out}} & = \mathcal{O}(\Delta^{-1}) \label{eq:lambda_out_asymp}.
\end{align}
Assuming charge neutrality, $Z_{\mathrm{tot}}=\mathcal{O}(\eta)$, so Eq.~\eqref{eqn:lambda_SOS} immediately gives
\begin{equation}
    \lambda_{\text{SOS}} = \lambda_T + \left((\eta + Z_{\text{tot}})^2 + \eta\right)\lambda_{\text{out}} = \mathcal{O}(\eta \Delta^{-2} + \eta^2 \Delta^{-1}).
    \label{eq:lambda_sos_asymp}
\end{equation}
The true ground-state energy $E_0$ can be upper bounded by any extensive, variational energy such as the Hartree-Fock energy:
\begin{equation}
    E_0 < E_{\mathrm{HF}} = \mathcal{O}(\eta).
\end{equation}
The extensivity of the HF energy is a trivial case of the linked cluster theorem (see Chap.~1 and 3 of Ref.~\cite{fetter2003quantum} and Chap.~5 of Ref.~\cite{mattuck1992guide}), which states that unlinked diagrams cancel out at each order in perturbation theory. This means that the exact ground state energy and approximate ground energies that are constructed from only linked diagrams are extensive. This theorem is expected to hold broadly for systems of particles with 2-body Coulomb interactions. We also assume that the Hartree-Fock energy will converge to a basis set limit,
\begin{equation}
    E_{\mathrm{HF}}(\Delta^{-1}) = E_{\mathrm{HF, BSL}} + f(\Delta^{-1}), \qquad
    \lim_{\Delta^{-1} \rightarrow \infty} f(\Delta^{-1}) = 0,
\end{equation}
and is therefore independent of $\Delta^{-1}$ asymptotically. 
As a result, $E_0$ should not have significant dependence on $\Delta^{-1}$, and we can give
\begin{equation}
    E_0 = \mathcal{O}(\eta).
\end{equation}

The Hamiltonian shift $\beta$ is given in Eq.~\eqref{eq:beta-def} as the sum of two terms, $\eta\lambda_{\text{out}}$ and $\beta_{\text{nuc}}$.
The term $\eta\lambda_{\text{out}}$ scales as $\mathcal{O}(\eta \Delta^{-1})$ due to Eq.~\eqref{eq:lambda_out_asymp}.
The scaling of $\beta_{\text{nuc}}$ is more easily considered by comparing to the non-Born Oppenheimer case.
In that case,
\begin{equation}
    \beta_{\text{nuc}}^{\rm non-BO} = \lambda_{\text{out}} \sum_{\ell}\zeta_{\ell}^2 \, ,
\end{equation}
which is similar to the first term $\eta\lambda_{\text{out}}$ for electrons which can be regarded as $\lambda_{\text{out}}$ times the sum of squares of electron charges.
In this case $\beta_{\text{nuc}}^{\rm non-BO}=\mathcal{O}(\eta)$ because the nuclear charges are $\mathcal{O}(1)$ and there are no more nuclei than electrons.

In the non-Born Oppenheimer case there is also the term in the Hamiltonian for the nuclear potential, but that should be $\mathcal{O}(\eta)$ for similar reasons as the electronic potential.
The way $\beta_{\text{nuc}}$ differs from $\beta_{\text{nuc}}^{\rm non-BO}$ is that it has an extra term corresponding to a classical approximation of the nuclear potential, 
\begin{align}\label{eq:beta-nuc-expanded}
     \beta_{\text{nuc}} &= \beta_{\text{nuc}}^{\rm non-BO} +
     \sum_{\nu \in G_0}C_\nu
      \sum_{\substack{\ell,\ell' \\ \ell\ne \ell'}}\zeta_{\ell}\zeta_{\ell'} e^{-ik_{\nu}\cdot (R_{\ell}-R_{\ell'})} \nn
      &\approx \beta_{\text{nuc}}^{\rm non-BO} +\sum_{\ell} \zeta_\ell  V_{\ell}( R_{\ell} )\, ,
\end{align}
where
\begin{align}\label{eq:Veff}
V_\ell (R_\ell) &= \sum_{\ell'\ne \ell} \zeta_{\ell'} v(R_{\ell} - R_{\ell'}) \, , \\
    v(R) &= \sum_{\mathbf{T}} \left[\frac{1}{2\|R + \mathbf{T}\|} - \frac 1\Omega\int_{\text{box }\mathbf{T}} \frac{1}{2\|R - r\|} \, d^3r \right] ,
\end{align}
with $\mathbf{T}$ being translations of the simulation box.
The effective potential $V_\ell$ corresponds to a periodic system with a uniform background charge.
The uniform compensating background charge for the nuclei cancels exactly the uniform compensating background charge for the electrons which allows us to separately omit the divergent $\nu =0$ in each term.
Because the effective potential is for a charge-neutral system, the total potential in $\beta_{\text{nuc}}$ is extensive and proportional to $\eta$.
This shows that $\beta_{\text{nuc}}=\mathcal{O}(\eta)$, similar to the non-Born Oppenheimer case.
See Appendix \ref{app:beta_scaling} for a more detailed explanation and numerical justification.

Substituting these scalings into Eq.~\eqref{eq:lambda_eff_bound} yields 
\begin{align}\label{eq:sossa_lambda_effective}
    \lambda_{\text{eff}}
    &= \mathcal{O}(\eta \Delta^{-1.5} + \eta^{1.5} \Delta^{-1}).
\end{align}
We can now compare directly to the standard 1-norm required by prior first-quantized LCU algorithms \cite{su2021fault,berry2024quantum}, which scales as $\lambda_{\text{LCU}}= \mathcal{O}(\eta \Delta^{-2} + \eta^2 \Delta^{-1})$ as stated in Eq.~\eqref{eqn:lambda_lcu_scaling}. This reveals two different asymptotic speedups depending on the target simulation regime:
\begin{enumerate}[label=\roman*)]
    \item Continuum Limit ($\Delta^{-1} \gg \eta$) where the kinetic energy dominates, and the algorithm complexity is reduced from $\mathcal{O}(\eta \Delta^{-2})$ to $\mathcal{O}(\eta \Delta^{-1.5})$, representing an $\mathcal{O}(\Delta^{-0.5})$ speedup,
    \item Large-System Limit ($\eta \gg \Delta^{-1}$) where the Coulomb interactions dominate, and the algorithm complexity is reduced from $\mathcal{O}(\eta^2 \Delta^{-1})$ to $\mathcal{O}(\eta^{1.5} \Delta^{-1})$, giving an $\mathcal{O}(\eta^{0.5})$ factor improvement.
\end{enumerate}
In the non-Born Oppenheimer case we obtain the same asymptotic scaling, because $\lambda_{\text{SOS}}$ is unchanged and the scaling of $\beta$ is unchanged.
Thus these speedups hold in both the Born Oppenheimer and non-Born Oppenheimer cases.

\section{Compilation and logical resource estimates}\label{sec:logical_resources}
\subsection{SOSSA block encoding summary}
Here we provide a high-level description of the block encoding for the walk operator used for ground state-energy estimation. Specific compilation details are provided in Appendix~\ref{app:block_encoding}. Given a generic SOS decomposition of a Hamiltonian like in Eq.~\eqref{eq:sos_h_sossa}
and block encodings $\textsc{Be}\left[O_{\alpha}/\lambda_{\alpha}\right]$ of $O_{\alpha}$  where
\begin{align}
    \textsc{Be}\left[O_{\alpha}/\lambda_{\alpha}\right] = 
    \begin{pmatrix}
        O_{\alpha}/\lambda_{\alpha} & \cdots \\
        \vdots & \ddots
    \end{pmatrix}
    \Rightarrow \Pi_{l}\textsc{Be}\left[O_{\alpha}/\lambda_{\alpha}\right]\Pi_{m} = O_{\alpha}/\lambda_{\alpha} \otimes |0\rangle\langle 0|_{\text{ancilla}}
\end{align}
such that $\{\Pi_{l},\Pi_{m}\}$ are orthogonal projectors of dimension $l$ and $m$ onto the zero-strings $|0\rangle_{l}$ and $|0\rangle_{m}$ 
then one can construct a block encoding of a single qubitization step as a sum over controlled block encodings $\sum_{\alpha}|\alpha\rangle\langle\alpha| \otimes \textsc{Be}\left[T_{2}\left(\textsc{Be}\left[O_{\alpha} / \lambda_{\alpha}\right] \right)\right]$ where $T_2(A)=2A^\dagger A-\mathbb{I}$ is a generalization of the Chebyshev polynomial for operators, such that
\begin{align}\label{eq:sossa_lcu_t2}
    \textsc{Be}\left[2 (H_{\text{SOSSA}}^{\dagger}H_{\text{SOSSA}} / \lambda_{\text{SOS}}) - 1\right] = \left(\textsc{PREP}^{\dagger}  \otimes \mathbb{I} \right) \left( \sum_{\alpha}|\alpha\rangle\langle\alpha| \otimes \textsc{Be}\left[T_{2}\left( \textsc{Be}\left[ \frac{O_{\alpha}}{\lambda_{\alpha}} \right] \right) \right] \right) \left(\textsc{PREP} \otimes\mathbb{I} \right)
\end{align}
where $\textsc{PREP}$ prepares an ancilla state as an appropriately weighted superposition over $\ket\alpha$.
A unitary walk operator $W$ can be constructed from the block encoding in Eq.~\eqref{eq:sossa_lcu_t2} by taking the product with a reflection unitary on the signal state $\ket\alpha$ and any signal states for the block encodings of $O_{\alpha}$,
\begin{equation}
    W \equiv (2\Pi_{\mathrm{all}} - \mathbb{I})\textsc{Be}\left[2 (H_{\text{SOSSA}}^{\dagger}H_{\text{SOSSA}} / \lambda_{\text{SOS}}) - 1\right].
\end{equation}
The walk operator has eigenphases given in Eq.~\eqref{eq:sossa_phases}.  The first-order error propagation formula can be used to determine the number of queries $N_{W}$ necessary to estimate the energy $E_{0}$ with error $\sigma_{E_{0}}$ \cite{lowsossa2025}
\begin{align}\label{eq:qpe_cost}
N_{W} = \frac{\pi \lambda_{\text{eff}}}{2\sigma_{E_{0}}} ,\qquad  \lambda_{\text{eff}} = \sqrt{E_{\text{gap}} \left( \lambda_{\text{SOS}} - E_{\text{gap}}  \right)} \, .
\end{align}
Note that a factor of $2$ difference under the square root in Eq.~\eqref{eq:qpe_cost} compared with Ref.~\cite{lowsossa2025} originates from our definition of $\lambda_{\text{eff}}$ differing by a factor of 1/2. 
Given SOS generators in Eqs.~\eqref{eq:t_sos_gen}, \eqref{eq:v_sos_gen}, and \eqref{eq:x_sos_gen} we use the following operator forms to generate each $\textsc{Be}\left[T_{2}\left( 
\textsc{Be}\left[ \frac{O_{\alpha}}{\lambda_{\alpha}} \right] \right) \right]$:
\begin{align}
    T_{j,w} &= \frac{1}{2}\sum_{b\in\{0,1\} }\sum_{r =0}^{n_p-2} \frac{\sqrt{2}\pi}{\Omega^{1/3}} 2^{r}  \sum_{p\in G} (-1)^{b(1+p_{w,r})} \ket{p}\bra{p}_j, \label{eq:t_sos_gen_be}\\
    U_{\text{shift},j}(\nu) &= \sum_{\substack{ p \in G: \\ p - \nu \in G}}\ket{p-\nu}\bra{p}_{j}, \\
    V_{\nu} &=  \sqrt{C_{\nu}}\tilde{V}_{\nu} \, ,\quad C_{\nu} = \frac{2\pi}{\Omega \|k_{\nu}\|^{2}} = \frac{1}{2\pi\Omega^{1/3} \|\nu\|^{2}} \, , \\
    \tilde{V}_{\nu} &= \sum_{j=1}^{\eta}U_{\text{shift},j}(\nu) - \sum_{\ell=1}^{L}\zeta_{\ell}e^{-i k_{\nu}\cdot R_{\ell}}\mathbb{I} \\
    &=\sum_{j=1}^{\eta}U_{\text{shift},j}(\nu) - \sum_{m=1}^{Z_{\text{tot}}}e^{-i k_{\nu}\cdot R_{\ell}(m)}\mathbb{I} \, , \label{eq:v_sos_gen_extended} \\
    X_{j,\nu}  &= \sqrt{C_{\nu}}\tilde X_{j,\nu} , \quad \tilde X_{j,\nu}=\sum_{\substack{p \in G: \\p-\nu \notin G}}  \ket{p-\nu} \bra{p}_j.
    \label{eq:x_sos_gen_be}
\end{align}

Before proceeding to a more detailed description of the implementation strategy, note that we expressed the sum over $\ell$ as the sum over total charge $Z_{\text{tot}}$ to emulate the $\zeta_{\ell}$ weighting from the LCU by repeating the same unitary $\zeta_{\ell}$ times.
To see this, consider the inner LCU of $V_{\nu}$ associated with applying operator $e^{-ik_{\nu}\cdot R_{\ell}}\mathbb{I}$. The signal state will be
\begin{align}
    |\psi_{\text{prep}}\rangle = \frac{1}{\sqrt{Z_{\text{tot}}}}\sum_{m=1}^{Z_{\text{tot}}}|m\rangle
\end{align}
and when combined with \textsc{SEL}
\begin{align}
    \textsc{SEL} = \sum_{m=1}^{Z_{\text{tot}}}|m\rangle\langle m| \otimes e^{-i k_{\nu}\cdot R_{\ell}(m)}\mathbb{I}
\end{align}
can be seen to produce a block encoding of our desired operator
\begin{align}
    \textsc{Be}\left[H_{\text{eff}} / Z_{\text{tot}}\right] &= \langle \psi_{\text{prep}}|\textsc{SEL}|\psi_{\text{prep}}\rangle \\
    &=  \sum_{m=1}^{Z_{\text{tot}}}\left(\frac{1}{\sqrt{Z_{\text{tot}}}}\right)^{2}e^{-i k_{\nu}\cdot R_{\ell}(m)}\mathbb{I} \\
    &= \sum_{\ell=1}^{L}\left(\sum_{m \mapsto \ell}\frac{1}{Z_{\text{tot}}}\right)e^{-i k_{\nu}\cdot R_{\ell}} \mathbb{I} \\
     &= \sum_{\ell=1}^{L}\left(\frac{\zeta_{\ell}}{Z_{\text{tot}}}\right)e^{-i k_{\nu}\cdot R_{\ell}} \mathbb{I} \, .
\end{align}
We perform this modification because we can use existing uniform superposition circuit preparations~\cite{babbush2018encoding} to minimize the number of quantum read-only memory (QROM) table lookup implementations. Furthermore, to save qubits we can combine the register indexing $m$, iterating over $Z_{\text{tot}}$, and register indexing the electrons $j$ because we never sum over both indices simultaneously.  This new index will be in a register of $n_{\texttt{idx}} = \lceil \log_{2}\left( \max\left[\eta, Z_{\text{tot}}\right]\right)\rceil$ bits.

With the SOS generators in the form of Eqs.~\eqref{eq:t_sos_gen_be}-\eqref{eq:x_sos_gen_be}, both the `inner' and `outer' block encodings in this nested structure are naturally implemented as LCUs. This means that we must implement an outer prepare (PREP) oracle that prepares a superposition over the $\alpha$ register which indexes terms in the SOS, an inner prepare that prepares a superposition over ancillae needed to block encode each $O_{\alpha}$, and a select (SEL) operator that applies the individual unitaries controlled on all the ancillae. Notice that both the momentum shift operator $U_{\text{shift}, j}(\nu)$ (within $V_{\nu}$) and the self-interaction operator $X_{j,\nu}$ rely on computing a momentum translation $p - \nu$, but they are restricted to mutually exclusive domains: $p - \nu \in G$ (in-bounds) and $p - \nu \notin G$ (out-of-bounds), respectively. 

We reduce the arithmetic overhead by absorbing this conditional bounds-checking directly into the structural reflection required to block encode the Chebyshev polynomial. Recall that constructing $\textsc{Be}\left[T_{2}\left(
\textsc{Be}\left[ \frac{O_{\alpha}}{\lambda_{\alpha}} \right] \right) \right]$ requires applying the forward block encoding, performing a reflection about the inner ancilla zero-state, and then applying the reverse block encoding~\cite{lowsossa2025, von2021quantum}. 
To exploit this without introducing complex conditional arithmetic, we allocate a single `overflow-flag' qubit, initialized to $|0\rangle$, within the inner ancilla space. During the forward block encoding unitary ($U = \textsc{Be}\left[O_{\alpha}/\lambda_{\alpha}\right]$) for both the $V_{\nu}$ and $X_{j,\nu}$ branches, the circuit computes the momentum shift $p - \nu$ and writes the boundary status to this flag ($|1\rangle$ if out-of-bounds, $|0\rangle$ if in-bounds). The inner reflection operator of the walk step then applies a $-1$ phase to any state orthogonal to the all-zero ancilla. For the $V_{\nu}$ branch, out-of-bounds states carry the $|1\rangle$ flag, ensuring they receive this phase flip and act as the block encoding garbage state. For the $X_{j,\nu}$ branch, we apply a NOT gate to the `overflow-flag' prior to the reflection; this logically inverts the target, routing in-bounds states to garbage and promoting the out-of-bounds states to the signal subspace. Finally, we apply the reverse block encoding ($U^\dagger$). This protocol places all appropriate states for $V_{\nu}$ and $X_{j,\nu}$ in the desired subspace and all others in the qubitization garbage state.

\subsection{Separation of outer and inner prepare}
In prior presentations of SOSSA for ground-state energy estimation \cite{lowsossa2025, king2026quantum} a clean separation between the $\alpha$ index and any indices needed for block encoding $O_{\alpha}$, inner and outer indices, was assumed. 
Notice that in Eq.~\eqref{eq:sos_decomposition} the electron summation $j$ is on the outside summation for the SOS generators associated with the kinetic energy operator and the self-interaction adjustment $X$, but on the inside of the SOS generator associated with the potential terms $V$. If our ancilla space hosting block encoding signal states consisted of
\begin{align}
    \mathcal{H}_{\text{ancilla}} = \mathcal{H}_{\nu} \otimes \mathcal{H}_{w} \otimes \mathcal{H}_{r} \otimes \mathcal{H}_{\texttt{idx}} \otimes \mathcal{H}_{\text{overflow}} \otimes \mathcal{H}_{b}
\end{align}
where each Hilbert space is hosting the \textsc{PREP} state for the LCU indexed indexed by the subscript, then there is no clean separation between outer indices and inner indices.
This is primarily because $\texttt{idx}$ indexes the outer sum over electrons for $T$ and $X$, but the inner sum over electrons and nuclei for $V_\nu$.

We could duplicate registers to achieve the separation but here we take a different approach to save qubits. To enable a clean separation between outer and inner summation indices, we can create a block-diagonal Hilbert space by augmenting with selection bits. We define a new register $S$ with state $s \in \{0, 1, 2\}$ and use the fact that the augmented ancilla space can be used to express a clean separation between outer and inner indices for each $s$ in a direct sum fashion
\begin{align}\label{eq:separated_outer_inner_direct-sum}
\mathcal{H}_{\text{ancilla}} = \bigoplus_{s=0}^{2}\left(|s\rangle\langle s|_{S} \otimes \mathcal{H}_{\text{out}}^{(s)} \otimes \mathcal{H}_{\text{in}}^{(s)} \right).
\end{align}

The three states in $S$ correspond to each term in the SOS we are block encoding
\begin{align}\label{eq:hilbertspace_trunc}
& s=0 \;(\text{Kinetic}\; T): \mathcal{H}_{\text{out}}^{(0)} = \mathcal{H}_{\texttt{idx}} \otimes \mathcal{H}_{w}\;,\;\;\mathcal{H}_{\text{in}}^{(0)} = \mathcal{H}_{b} \otimes \mathcal{H}_{r} \\
& s=1 \;(\text{Potential} \; \tilde{V}_{\nu}): \mathcal{H}_{\text{out}}^{(1)} = \mathcal{H}_{\nu}\;,  \;\; \mathcal{H}_{\text{in}}^{(1)} = \mathcal{H}_{f_{\text{nuc-flag}}}\otimes \mathcal{H}_{\texttt{idx}} \otimes \mathcal{H}_{\text{overflow}}  \\
& s=2 \;(\text{Self-interaction}\; X): \mathcal{H}_{\text{out}}^{(2)} = \mathcal{H}_{\texttt{idx}} \otimes \mathcal{H}_{\nu}\;,\;\; \mathcal{H}_{\text{in}}^{(2)} = \mathcal{H}_{\text{overflow}} .
\end{align}
An appropriate weighting of each of the three subspaces must be integrated into the outer \textsc{PREP} of the block encoding. This begins with a branch selection operator $P_{\text{out,branch}}$ acting on the $S$ register, creating a superposition over the three branches weighted by the total 1-norm of each respective Hamiltonian component:
\begin{align}
    P_{\text{out,branch}}|0\rangle_{S} &= \sqrt{\frac{\Lambda_{T}}{\lambda_{SOS}}} |0\rangle_{S}|\phi_{\text{out}}^{(0)}\rangle + \sqrt{\frac{\Lambda_{V}}{\lambda_{SOS}}} |1\rangle_{S} |\phi_{\text{out}}^{(1)}\rangle + \sqrt{\frac{\Lambda_{X}}{\lambda_{SOS}}} |2\rangle_S |\phi_{\text{out}}^{(2)}\rangle,
\end{align}
where $\Lambda_T$, $\Lambda_V,$ and $\Lambda_X$ are the respective 1-norm contributions from the kinetic, potential, and self-interaction terms (e.g., $\Lambda_T = \sum_{j,w} \lambda_{T_{j,w}}^2$) and $|\phi_{\text{out}}^{(s)}\rangle$ is the superposition over $|\alpha\rangle$ for this component of the SOS. Once this branch superposition is established, the subsequent inner preparation operators $P_{\text{in}}$ are applied as block-diagonal unitaries strictly controlled by $S$. 

With the direct-sum space established, we can now verify core criteria of the SOSSA block encoding protocol in the extended ancilla space. Specifically, we must show that the inner LCU correctly projects to the desired operator for each branch. Let the inner state preparation unitary be defined block-diagonally over our direct-sum space as 
$P_{\text{in}} = \sum_{s=0}^2 |s\rangle\langle s|_{S} \otimes P_{\text{in}}^{(s)}$. 
Applying this to the $|0\rangle_{s}$ inner states (the ancilla space for each inner Hilbert space in Eq.~\eqref{eq:hilbertspace_trunc}) generates the inner signal states for each branch, 
$|\phi_{\text{in}}^{(s)}\rangle = P_{\text{in}}^{(s)}|0\rangle_{\text{in}}^{(s)}$. 
Similarly, the \textsc{SEL} oracle is branch-dependent, with
$S_{\text{in}} = \sum_{s=0}^2 |s\rangle\langle s|_{S} \otimes S_{\text{in}}^{(s)}$, where $S_{\text{in}}^{(s)}$ acts upon $|\phi_{\text{out}}^{(s)}\rangle$ as well to give the required dependence on the outer summation indices. To demonstrate the criterion, 
the block encoding relies on the projection of the inner registers back to their initial state. In our direct-sum space, this projection operator is defined as:
\begin{align}
\Pi_{\text{in}} &= \sum_{s=0}^{2} |s\rangle\langle s|_{S} \otimes \mathbb{I}_{\text{out}}^{(s)} \otimes |0\rangle\langle 0|_{\text{in}}^{(s)}
\end{align}
where the identity operator changes in size appropriately to match the size of the inner register. We must verify the expectation value of the inner LCU, evaluating $\langle 0|_{\text{in}} P_{\text{in}}^\dagger S_{\text{in}} P_{\text{in}} |0\rangle_{\text{in}}$. Producing the correct terms relies on the branch register states being orthogonal ($\langle s | s' \rangle = \delta_{s,s'}$). Thus, we obtain inner LCUs:
\begin{align}
\langle 0|_{\text{in}} \left( P_{\text{in}}^\dagger S_{\text{in}} P_{\text{in}} \right) |0\rangle_{\text{in}} &= \sum_{s=0}^2 |s\rangle\langle s|_{S} \otimes  \left( \langle \phi_{\text{in}}^{(s)} | S_{\text{in}}^{(s)} | \phi_{\text{in}}^{(s)} \rangle \right).
\end{align}
Each branch evaluates to its respective normalized operator from the SOS decomposition
\begin{align}
\langle \phi_{\text{in}}^{(0)} | S_{\text{in}}^{(0)} | \phi_{\text{in}}^{(0)} \rangle &= \sum_{j,w}\ket{j}\bra{j}\otimes\ket{w}\bra{w} \otimes \frac{T_{j,w}}{\lambda_{\text{in}}^{(0)}}, \\
\langle \phi_{\text{in}}^{(1)} | S_{\text{in}}^{(1)} | \phi_{\text{in}}^{(1)} \rangle &= \sum_\nu \ket{\nu}\bra\nu \otimes \frac{\tilde{V}_{\nu}}{\lambda_{\text{in}}^{(1)}} \label{eq:inner_block_encoding_V}, \\
\langle \phi_{\text{in}}^{(2)} | S_{\text{in}}^{(2)} | \phi_{\text{in}}^{(2)} \rangle &= \sum_{j,\nu} \ket{j}\bra{j}\otimes\ket{\nu}\bra\nu \otimes\frac{X_{j,\nu}}{\lambda_{\text{in}}^{(2)}},
\end{align}
giving the correct block-encoded operator. We now discuss the implementation strategy and specific optimizations.

\subsection{Circuit Implementation}\label{sec:circuit_implementation}
A circuit diagram showing the schematic structure of the block encoding is shown in Fig.~\ref{fig:circuit}, and more details are provided in Appendix~\ref{app:block_encoding}.
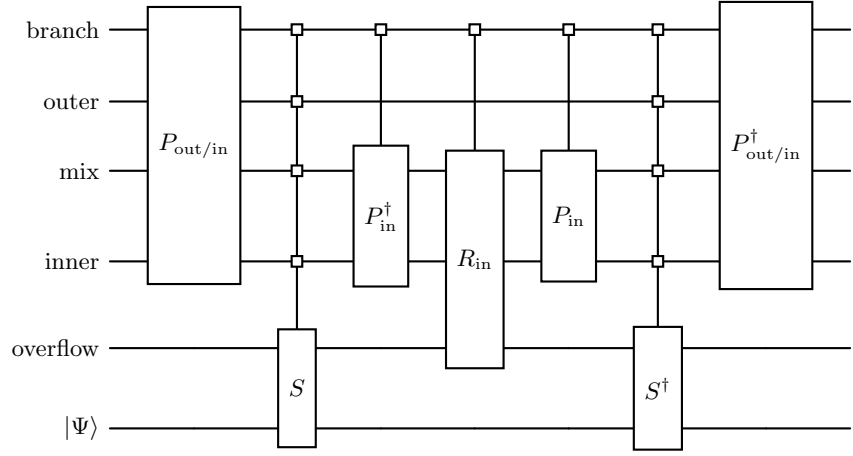
\begin{figure}
    \centering
    \begin{quantikz}
    \lstick{branch}
        & \gate[4]{P_{\text{out/in}}}
        & |[operator]| 
        & |[operator]| 
        & |[operator]| 
        & |[operator]|
        & |[operator]|
        & \gate[4]{P_{\text{out/in}}^{\dagger}}
        &
        \\
    \lstick{outer}
        & 
        &|[operator]| 
        &
        &
        &
        &|[operator]| 
        &
        &
        \\
    \lstick{mix}
        & 
        &|[operator]| 
        & \gate[2]{P_{\mathrm{in}}^{\dagger}}\vqw{-2}
        & \gate[3]{R_{\mathrm{in}}}\vqw{-2}
        & \gate[2]{P_{\mathrm{in}}}\vqw{-2}
        & |[operator]| 
        &
        &
        \\
    \lstick{inner}
        & 
        &|[operator]| 
        &
        &
        &
        &|[operator]| 
        &
        &
        \\
    \lstick{overflow}
        & 
        & \gate[2]{S}\vqw{-4}
        &
        &
        &
        & \gate[2]{S^{\dagger}}\vqw{-4}
        &
        &
        \\
    \lstick{$\ket{\Psi}$}
        &
        & 
        &
        &
        &
        &
        &
        &
        \\
    \end{quantikz}
    \caption{Overall structure of the circuit for the block-encoding. For clarity we have grouped the registers as follows: ``branch" indicates the qubits used to control the application of the $T$, $V$, and $X$ branches, ``outer" indicates those qubits involved only in the outer state preparation (nested boxes and the $w$ register), ``mix" indicates the shared index register and nuclear flag which are part of the outer or inner prepare depending on which term is being applied, ``inner" indices qubits involved only in the inner state preparation (the $r$ register and the $b$ flag), ``overflow" indicates overflow bits, and $\ket{\Psi}$ is the system register. We use a small square to indicate some kind of generalized control.}
    \label{fig:circuit}
\end{figure}

\subsubsection{Preparation oracles}
Utilizing the direct sum decomposition we can now construct the states and unitaries needed for the  block encoding described in Eq.~\eqref{eq:sossa_lcu_t2}. The states we need to prepare are
\begin{align}
    |\Psi_{\text{prep}}\rangle &= P_{\text{out/in}} |0\dots 0\rangle_{\text{anc}} = \bigoplus_{s=0}^{2} \sqrt{\frac{\Lambda_{(s)}}{\lambda_{\text{SOS}}}} |s\rangle_{S} \otimes |\phi_{\text{out}}^{(s)}\rangle \otimes |\phi_{\text{in}}^{(s)}\rangle, \label{eq:final_prep_state}
\end{align}
where
\begin{align}
     |\phi_{\text{out}}^{(0)}\rangle &= \left( \frac{1}{\sqrt{\eta}} \sum_{j=1}^{\eta} |j\rangle_{\texttt{idx}} \right) \otimes \left( \frac{1}{\sqrt{3}} \sum_{w=0}^{2} |w\rangle \right), \\
     |\phi_{\text{in}}^{(0)}\rangle &= |+\rangle_b \otimes \left( \sum_{r=0}^{n_p-2} \sqrt{\frac{2^r}{2^{n_p-1}-1}} |r\rangle \right), \\
  |\phi_{\text{out}}^{(1)}\rangle &= \frac 1{\sqrt{\lambda_\nu}}\sum_{\nu \in G_0} \frac{1}{\|\nu\|} |\nu\rangle, \\
  |\phi_{\text{in}}^{(1)}\rangle &= \sqrt{\frac{\eta}{\eta + Z_{\text{tot}}}} |0\rangle_{f_{\text{nuc-flag}}} \left( \frac{1}{\sqrt{\eta}} \sum_{j=1}^{\eta} |j\rangle_{\texttt{idx}}\right) + \sqrt{\frac{Z_{\text{tot}}}{\eta + Z_{\text{tot}}}} |1\rangle_{f_{\text{nuc-flag}}} \left( \sum_{m=1}^{Z_{\text{tot}}} \sqrt{\frac{1}{Z_{\text{tot}}}} |m\rangle_{\texttt{idx}}  \right), \\
    |\phi_{\text{out}}^{(2)}\rangle &= \left( \frac{1}{\sqrt{\eta}} \sum_{j=1}^{\eta} |j\rangle_{\texttt{idx}} \right) \otimes \left( \sum_{\nu \in G_0} \frac{1}{\|\nu\| \sqrt{\lambda_\nu}} |\nu\rangle \right).
\end{align}
The state preparation for the $|\nu\rangle$ register is always in the outer \textsc{PREP} operation and thus we can use the same probability recycling introduced in Ref.~\cite{su2021fault, babbush2019quantum} for implementing the two potential block encodings for $V_{\nu}$ and $X_{j,\nu}$.
The nested boxes state preparation for the outer state over the momentum difference register $\nu$ can be applied either with or without amplitude amplification.
In the case it is without amplitude amplification, it has a success probability $p_{\nu}$ of approximately $1/4$, so substantial probability recycling is used.
In the case with amplitude amplification, a single round boosts the success probability to near one at the expense of a longer circuit (approximately three times longer).
This trade-off means that it may be optimal to use either amplitude amplification or just probability recycling depending on the parameters of the system.
We provide a full description of the modifications needed for SOSSA in Appendix~\ref{app:be_nu_state_prep}. Furthermore, we also reuse the preparation routines from Ref.~\cite{su2021fault} for $\{w,j,r\}$.

To represent the $S$ register, which holds three values, we use two ancilla qubits, $q_T$ and $q_{VX}$. The qubit $q_{VX}$ is rotated by an angle $\varphi$ to divide the success amplitude exactly between $V$ and $X$. Because both $V$ and $X$ require the $|\nu\rangle$ register and are subject to the same failure rate $p_\nu$, their relative phase is defined as: 
\begin{align}
    \cos^2\varphi = \frac{\Lambda_V}{\Lambda_V + \Lambda_X}, \quad \sin^2\varphi = \frac{\Lambda_X}{\Lambda_V + \Lambda_X}.
\end{align}

The qubit $q_T$ is rotated by an angle $\theta$ to balance $T$ against the $(V+X)$ subsystem. The angle is set according to the discussion in Appendix~\ref{app:be_nu_state_prep}. With knowledge of how to set $\theta$ and $\varphi$ we allocate two ancilla qubits, $q_T$ and $q_{VX}$, initialized to $|00\rangle$. Using the synthesis bounds established in Ref.~\cite{su2021fault}, an arbitrary single-qubit rotation synthesized to precision $\epsilon_T$ requires $n_T = \lceil \log_2(\pi \lambda / \epsilon_T) \rceil$ bits of precision. The nested boxes subroutine is executed (uncontrolled) to prepare the $|\nu\rangle$ register, yielding the failure flag $f_\nu$. To minimize control overhead inside the $\mathcal{O}(\eta)$ \textsc{SEL} oracle, we resolve the tuple $(q_T, q_{VX}, f_\nu)$ into three distinct, mutually exclusive execution flags ($\text{Exec}_T, \text{Exec}_V, \text{Exec}_X$). Details about circuit construction for these flags can be found in Appendix~\ref{app:exec_flag_calc}.

The remaining component of the \text{PREP} circuits is the controlled preparation over $j$ and $Z_{\text{tot}}$ for the outer summation in $T$ and $X$ and the inner summation in $V$. The only non-trivial inner preparation is for the $V_{\nu}$ term which has two LCUs involving index $j$ and index $m$.
Because there are two terms, we prepare a flag qubit $f_\text{nuc-flag}$ to split the amplitude between $j$ for the electron-electron interaction ($|0\rangle$) and $m$ for the electron-nuclear interaction ($|1\rangle$). Because this split is only required by the potential operator, the rotation is controlled on $\text{Exec}_V = |1\rangle$. For the $T$ and $X$ branches, this operation is bypassed, and $f_{\text{nuc-flag}}$ naturally remains $|0\rangle$.

For a simulation involving a charge neutral system ($\eta = Z_{\text{tot}}$) the same uniform superposition would be needed for all terms and thus can be uncontrolled.  We would still need to ensure that we appropriately apply the reflection around a $|0\rangle_{\texttt{idx}}$ state when simulating $\textsc{Be}\left[T_{2}\left(\textsc{Be}\left[V_{\nu}/\lambda_{V_{\nu}}\right]\right)\right]$ (inner reflection).  To accomplish this, having just performed an uncontrolled preparation on register \texttt{idx} we would then need to unprepare before reflecting.  We can use the fact that the reflection can be controlled based on $\text{Exec}_{V}$ such that identity is applied for the $X$ and $T$ branches, but the reflection is applied for the $V$ branch. 
That is, for the central reflection of the block encoding used to apply the oblivious amplitude amplification, the reflection would include the qubits where $j$ and $m$ are prepared for the $V$ branch, but not for the $X$ and $T$ branches.
Therefore, the total Toffoli cost for the $\texttt{idx}$ preparation with a charge-neutral system is
\begin{align}
    \text{cost-neutral} = 4  (3 n_{\eta} + 2b_{r} - 9)
\end{align}
where the factor of 4 comes from having to unprepare and prepare on the inner branch when performing the reflection for the $V$ block encoding. For non-charge neutral systems the cost grows a small amount. A detailed discussion on how to handle this case in Appendix~\ref{app:prep_inout_be}. The same reasoning applies to the preparation over the $r$ index and because the index only appears in the kinetic energy term its preparation and unpreparation can be uncontrolled. As a simple summary of all parts of the state-preparation procedure, we tabulate the merged $P_{\text{out/in}}$ in Table~\ref{tab:prep_costs}.

\begin{table}[H]
\scriptsize
\caption{Comprehensive Toffoli costs for the merged state preparation ($P_{\text{out/in}}$). Costs are explicitly divided into forward preparation and uncomputation (inverse preparation) for a single call. The total cost per walk operator is the sum of the forward and uncompute costs, multiplied by the number of calls within the walk operator.
}
\label{tab:prep_costs}
\begin{ruledtabular}
\begin{tabular}{l c c c c}
\textbf{Subroutine} & $P_{\text{out}}$ & $P_{\text{out}}^{\dagger}$ & \textbf{Total (1 Call)} & \textbf{Calls / $W$} \\
\hline
Spatial dimensions ($w$) & $13$ & $13$ & $26$ & $1$ \\
Branch prep rotations ($q_{T}, q_{VX}$) & $2n_T - 6$ & $2n_T - 6$ & $4n_T - 12$ & $1$ \\
Nested Boxes ($|\nu\rangle$, $f_\nu$) &  $3  n_p^2 + 11 n_p + 2+ 4 n_{\mathcal{M}}  (n_p + 1)$ &  $0$ (Measurement) & $3n_p^2 + 11n_p +2 + 4n_{\mathcal{M}}(n_p + 1)$ & $1$ \\
Semantic Logic ($\text{Exec}_{T, V, X}$) & $2$ & $0$ (Measurement) & $2$ & $1$ \\
\hline
\textbf{Subroutine} & $P_{\text{in}}$ & $P_{\text{in}}^{\dagger}$ & \textbf{Total (1 Call)} & \textbf{Calls / $W$} \\
\hline
Bit-index cascade ($r$) & $n_p - 2$ & $0$ (Measurement) & $n_p - 2$ & $2$ \\
Nuclear flag split ($f_{\text{nuc-flag}}$) & $n_T - 2$ & $n_T - 2$ & $2n_T - 4$ & $2$ \\
Electron / Nuclear uniform $\eta \ne Z_{\text{tot}}$ $(j)$ & $3n_\eta + 2b_r - 9\ldots$ & $3n_\eta + 2b_r - 9\ldots$ & $6n_\eta + 4b_r - 18\ldots$ & $2$ \\
 & ${} \ldots+ 2|n_\eta-n_{Z_{\text{tot}}}|+4$ & ${} \ldots+ 2|n_\eta-n_{Z_{\text{tot}}}|+4$ & ${} ...+ 4|n_\eta-n_{Z_{\text{tot}}}|+8$ & \\
Electron / Nuclear uniform $\eta = Z_{\text{tot}}$ $(j)$  & $3n_\eta + 2b_r - 9$ & $3n_\eta + 2b_r - 9$ & $6n_\eta + 4b_r - 18$ & $2$ \\
\end{tabular}
\end{ruledtabular}
\end{table}

\subsubsection{Selection oracles}
The \textsc{SEL} oracle proceeds in three distinct phases: (A) unary iteration over index $j=m \in [0, \max[\eta, Z_{\text{tot}}] - 1]$ where for each value we swap electron $j$ into a designated workspace register and  write $R_{\ell}(m)$ to an ancilla register, (B) the conditional Hamiltonian arithmetic controlled by the execution flags ($\text{Exec}_T, \text{Exec}_V, \text{Exec}_X$), and (C) uncomputation of step (A).

 The unary iteration step requires $\max[\eta, Z_{\text{tot}}] - 2$ Toffoli gates. At each step of the unary iteration (up to $\eta$) we require $3n_{p}$ Toffolis for swapping electron $j$ to the workspace. $R_{\ell}$ is written to an ancilla workspace which requires only Clifford operations. This is uncomputed at the same cost for step (C). The looping and swapping is called four times because the \textsc{SEL} oracle is called twice per walk operator call. These swap operations match the original LCU approach of Ref.~\cite{su2021fault} and are the dominant cost of the block  (See Figure~\ref{fig:subroutinecosts}).

 The kinetic energy, potential, and self-interaction correction selection operations all act on the workspace and utilize slightly modified forms of the primitives from Ref.~\cite{su2021fault}. The kinetic energy selection operation still employs the sum over binary of the magnitude part of the signed integer representation of the state, except here it is just the absolute value rather than the square, and the square is obtained via the step of oblivious amplitude amplification.
 We require a minor modification of the Ref.~\cite{su2021fault} algorithm to accommodate the execution flag $|\text{Exec}_{T}\rangle$. 

The potential term has two components: 1) a momentum shift and 2) phasing. The unitary $S_{\text{in}}$ acts on $|\nu\rangle \otimes |f_{\text{nuc-flag}}, \texttt{idx}, \text{overflow}\rangle \otimes  |\Psi\rangle$ as
\begin{align}
S_{\text{in}}^{(1)} = \sum_\nu \ket\nu\bra\nu \otimes  \left(\vphantom{\sum_{j=1}^\eta}\right.\underbrace{\sum_{j=1}^\eta |0, j\rangle\langle 0, j|_{\text{nuc-flag},\texttt{idx}} \otimes U_{\text{shift}, j}(\nu)}_{\text{Electron branch}} \;\; + \;\; \underbrace{\sum_{m=1}^{Z_{\text{tot}}} |1, m\rangle\langle 1, m|_{\text{nuc-flag},\texttt{idx}} \otimes \Big(- e^{-i k_\nu \cdot R_{\ell}(m)} \mathbb{I}\Big)}_{\text{Nuclear branch}} \left.\vphantom{\sum_{j=1}^\eta}\right).
\end{align}
The two branches will be controlled by the controlled population of an ancilla register $\tilde{\nu}$ depending on $f_{\text{nuc-flag}}$.  For the first term involving the shift, we populate $\tilde{\nu}$ by controlled swap from $\nu$ to $\tilde{\nu}$ controlled by $f_{\text{nuc-flag}}$. For the second term involving the nuclear phasing, we use the $\nu$ register and $|R_{\ell}\rangle$ for the multiply-accumulate circuit imparting the $e^{-ik_{\nu}\cdot R_{\ell}}$ phase (see Ref.~\cite[p.~20]{su2021fault}).

Note that we do not want the $e^{-ik_{\nu}\cdot R_{\ell}}$ phase in the branch for the kinetic energy $T$, but the approach of controlling this phase by the controlled swap of $\nu$ would result in this phase being applied with $S_{\text{in}}$.
However, because there is an outer sum over $j$ for $T$, applying $S_{\text{in}}^{\dagger}$ in the walk operator ensures that this phase is canceled, and no additional control logic is needed.
We use overflow ancilla bits as a component of the block encoding signal state to correctly block encode $V$ and $X$.
These are used to flag the $\nu$-shifted momentum being inside or outside $G_0$, as well as to ensure that $X^\dagger$ shifts back to the original momentum.
We include the relevant branching logic in Table~\ref{tab:branch_routing} for clarity.

\begin{table}[H]
\caption{\label{tab:branch_routing} Summary of data routing, momentum shifting, and phase accumulation across the four branches of the Hamiltonian. The controlled swap logic evaluates $\text{Exec}_X \lor (\text{Exec}_V \land \neg f_{\text{nuc-flag}})$ to determine whether the momentum transfer data resides in the active register $\tilde{\nu}$ or remains in $\nu$. Unwanted phases on the $T$ branch commute with the inner reflection and are canceled during the reverse walk operator step.}
\begin{ruledtabular}
\begin{tabular}{l c l l l}
\textbf{Active Branch} & \textbf{Control Logic} & \textbf{Data Location} & \textbf{Subtraction (reads $\tilde{\nu}$)} & \textbf{Phasing (reads $\nu$)} \\
\colrule
$V$ (Electron, $f_{\text{nuc-flag}}=0$) & True  & Swaps to $\tilde{\nu}$ ($\nu=0$) & Applies shift ($p - \tilde{\nu}$) & Identity ($e^0 = 1$) \\
$V$ (Nuclear, $f_{\text{nuc-flag}}=1$)  & False & Stays in $\nu$ ($\tilde{\nu}=0$) & Identity (subtracts 0) & Applies phase ($e^{-ik_\nu \cdot R_\ell}$) \\
$X$ (Self-Interaction)                  & True  & Swaps to $\tilde{\nu}$ ($\nu=0$) & Applies shift ($p - \tilde{\nu}$) & Identity ($e^0 = 1$) \\
$T$ (Kinetic)                           & False & Stays in $\nu$ ($\tilde{\nu}=0$) & Identity (subtracts 0) & Applies phase (Annihilated later) \\
\end{tabular}
\end{ruledtabular}
\end{table}

The $X_{j,\nu}$ term is largely block encoded already. 
In the previous step for $V$, we computed $(p-\nu)$ into the workspace register and converted it back to sign-magnitude form. Thus, we know if it overflowed the bounding box. We already computed a flag bit indicating if the subtraction overflowed so for $|\text{Exec}_{X}\rangle = |1\rangle$ we must flip the flag bit before reflecting. There is no additional non-Clifford cost specific to the $X$ branch because we flip the additional flag bit controlled by $\text{Exec}_{X}$ which is a CNOT. The total costs for all selection operations of $S_\text{in}$ and $S_{\text{in}}^{\dagger}$ are provided in Table~\ref{tab:select_costs}.
\begin{table}[H]
\footnotesize
\caption{\textbf{Comprehensive Toffoli Costs for the SELECT Oracle ($S_{\text{in}}$).}}
\label{tab:select_costs}
\begin{ruledtabular}
\begin{tabular}{l c c}
\textbf{Subroutine} & \textbf{Total (per $S_{\text{in}}$)} & \textbf{Calls / walk} \\
QROM + unary $\max[\eta, Z_{\text{tot}}]$ & $2\max[\eta, Z_{\text{tot}}] - 4$ & $2$ \\
CSWAP data to workspace & $6\eta n_p$ & $2$ \\
\textbf{Kinetic ($E_T$)} & & \\
Extract $p_{w,r}$ and apply CCZ phase & $3(n_p - 1)$ & $2$ \\
\textbf{Potential terms ($X$ and $V$)} & & \\
Subtraction ($\ket{p_{\text{work}} - \nu}$) & $12n_p + 1$ & $2$ \\
Phase Accumulation ($e^{-ik_\nu \cdot R_\ell}$) & $3  (2 n_p n_R - n_p (n_p + 1) - 1)$ & $2$ \\
Checking for overflow in $V$ and $X$ & $2$ & $2$ \\
\end{tabular}
\end{ruledtabular}
\end{table}
\subsubsection{Reflection costs}
The qubits we reflect on are described in Appendix~\ref{app:ref-section}. Here we provide the total costs in Table~\ref{tab:reflect_costs}.
\begin{table}[H]
\caption{Comprehensive Toffoli costs for the walk operator reflections and control for phase estimation. For the global reflection, in each step of phase estimation there is one more Toffoli to make this reflection controlled, as well as one more Toffoli for the unary iteration on the control register. For the inner reflection, the cost is the number of qubits minus 1, because an extra Toffoli is needed for control of the reflection on \texttt{idx}.
We define $n_{\text{idx}}\equiv\lceil \log_2 \max(\eta, Z_{\text{tot}}) \rceil$.}
\label{tab:reflect_costs}
\begin{ruledtabular}
\begin{tabular}{l c c}
\textbf{Reflection Subroutine} & \textbf{Total Toffoli Cost} & \textbf{Calls per Walk Op} \\
\hline
\multicolumn{3}{c}{\textit{Inner Reflection $(2\Pi_{\text{in}} - \mathbb{I})$}} \\
\hline \rule{0pt}{3ex}
\textbf{Kinetic ($E_T$):} Reflect on $r$ and $b$ & $n_p$ & $1$ \\
\textbf{Potential ($E_V$):} Reflect on $\texttt{idx}$, $f_{\text{nuc-flag}}$, \text{overflow-flag} & $n_{\text{idx}} + 2$ & $1$ \\
\textbf{Self-Interaction ($E_X$):} 0 & 0 & $1$ \\
\hline \rule{0pt}{3ex}
\textit{Subtotal for Inner Reflection:} & $n_p + n_{\text{idx}} + 1$ & \textit{1 (Sequential Sum)} \\
\hline
\multicolumn{3}{c}{\textit{Global Outer Reflection $(2\Pi_{\text{all}} - \mathbb{I})$}} \\
\hline \rule{0pt}{3ex}
Reflect on all active ancilla registers ($N_{\text{all}}$ qubits) & $4n_p + n_{\mathcal{M}} + n_{\text{idx}} + 14$ & $1$ \\
\end{tabular}
\end{ruledtabular}
\end{table}

\subsection{Gate costs of SOSSA}\label{sec:gate_costs}
Given the implementation described in Sec.~\ref{sec:circuit_implementation}, we show the logical gate costs of the end-to-end algorithm in Table~\ref{tab:comparison}. The gate counts depend on slightly inflated $\lambda_{\text{eff}}$ based on various implementation choices for nested-boxes and uniform state preparation that were described previously in Ref.~\cite{su2021fault}. The specific details for computing the inflated 1-norms are included in Appendix~\ref{app:inflated_1norm}. The $E_{\mathrm{gap}}$ that enters the equation for $\lambda_{\mathrm{eff}}$ (Eq.~\eqref{eq:lambda_eff}) is computed from $\beta$ (Eq.~\eqref{eq:beta-def}) and an upper bound on the ground-state energy, $E_0$. The upper bound on the true ground state electronic energy is obtained from the full, uncorrected Hartree-Fock energy without the (Ewald) energy of the nuclear lattice. In the case of the deuterium target from Ref.~\cite{rubin2024quantum}, the Hartree-Fock calculation proved too costly, so a slightly looser Hartree upper bound is used. These calculations were performed with Quantum ESPRESSO \cite{giannozzi2009quantum,giannozzi2017advanced}, and the ld1.x program was used to generate upf files for the bare Coulomb potential.

In Table~\ref{tab:comparison} we directly compare the first-quantized SOSSA algorithm with the LCU algorithm described in Ref.~\cite{su2021fault}. To benchmark the costs we used the two molecular systems from Ref.~\cite{su2021fault} (ethylene carbonate and LiPF$_6$), a set of supercells of lithium metal, and three of the targets (hydrogen, deuterium, and carbon) from Ref.~\cite{rubin2024quantum}. In all cases the gate cost of the block encoding is comparable to the LCU method which indicates that the SOSSA framework introduces minimal overhead. The reduction in the effective subnormalization constant leads to a speedup over the LCU method in all the systems shown here, and this speedup is particularly striking for the largest and most realistic systems (the hydrogen and deuterium targets from Ref.~\cite{rubin2024quantum}). Furthermore, since first-quantized SOSSA provides an asymptotic speedup, the speedup over the LCU method should only get bigger for larger systems. 

\begin{figure}[H]
    \centering
    \includegraphics[width=0.85\linewidth]{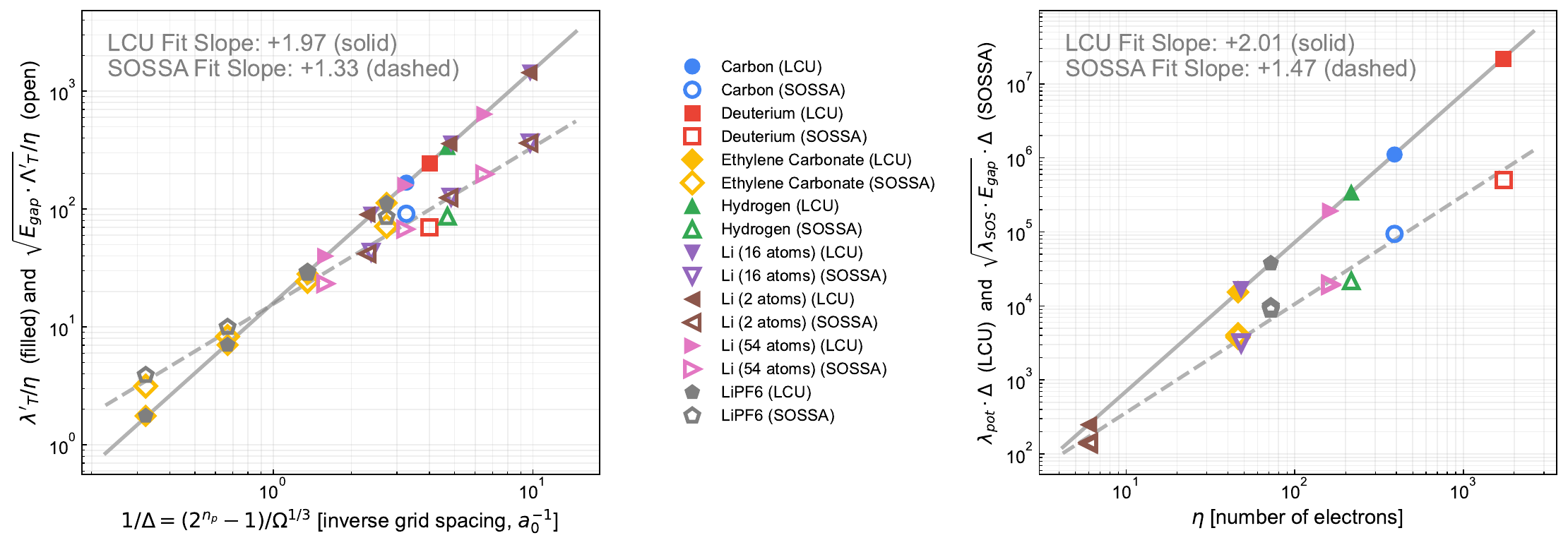}
    \caption{$\Lambda_T$ and $\Lambda_{\text{potential}}$ for LCU method from \cite{su2021fault} vs SOSSA (effective scaling after gap amplification). The lower slopes for SOSSA recover the expected scaling from Eq.~\eqref{eq:sossa_lambda_effective}.}
    \label{fig:lambda_T_VX_scaling}
\end{figure}

\begin{table}[H]
\centering
\caption{LCU~\cite{su2021fault} vs SOSSA: Side-by-Side Resource Comparison. Chemical accuracy target is $\varepsilon = 0.0016$ Hartree. The columns `LCU AA' and `SOS AA' indicate if amplitude amplification is applied to reduce the total gate complexity when considering different strategies for the $1/\|\nu\|$ state preparation.  More details about this tradeoff can be found in Appendix~\ref{app:be_nu_state_prep}.}
\label{tab:comparison}
\resizebox{\textwidth}{!}{
\begin{tabular}{l|rc|rrc|rr|rr|rr|r}
\hline\hline
\multicolumn{1}{c|}{\multirow{2}{*}{System}} & 
\multicolumn{1}{c}{\multirow{2}{*}{$\lambda_{\text{LCU}}$}} & 
\multicolumn{1}{c|}{\multirow{2}{*}{LCU AA}} & 
\multicolumn{1}{c}{\multirow{2}{*}{$\lambda_{\text{SOS}}$}} & 
\multicolumn{1}{c}{\multirow{2}{*}{$\lambda_{\text{eff}}$}} & 
\multicolumn{1}{c|}{\multirow{2}{*}{SOS AA}} & 
\multicolumn{1}{c}{HF} & 
\multicolumn{1}{c|}{\multirow{2}{*}{$-\beta$}} & 
\multicolumn{1}{c}{LCU} & 
\multicolumn{1}{c|}{SOS} & 
\multicolumn{1}{c}{LCU} & 
\multicolumn{1}{c|}{SOS} & 
\multicolumn{1}{c}{\multirow{2}{*}{Speedup}} \\
\multicolumn{1}{c|}{} & 
\multicolumn{1}{c}{} & 
\multicolumn{1}{c|}{} & 
\multicolumn{1}{c}{} & 
\multicolumn{1}{c}{} & 
\multicolumn{1}{c|}{} & 
\multicolumn{1}{c}{(no Ewald)} & 
\multicolumn{1}{c|}{} & 
\multicolumn{1}{c}{Walk} & 
\multicolumn{1}{c|}{Walk} & 
\multicolumn{1}{c}{QPE} & 
\multicolumn{1}{c|}{QPE} & 
\multicolumn{1}{c}{} \\
\hline
    ethylene carbonate (15 eV) & 5057.6 & AA & 29961.1 & 1300.3 & AA & -181.9 & -441.1 & 5,373 & 5,888 & $2.67 \times 10^{10}$ & $7.52 \times 10^{9\phantom{0}}$ & 3.5x \\
ethylene carbonate (60 eV) & 10636.4 & AA & 59922.2 & 2488.3 & AA & -289.8 & -739.8 & 6,693 & 7,368 & $6.99 \times 10^{10}$ & $1.80 \times 10^{10}$ & 3.9x \\
ethylene carbonate (250 eV) & 22283.3 & AA & 119844.4 & 5259.1 & AA & -361.9 & -1329.4 & 8,115 & 8,998 & $1.78 \times 10^{11}$ & $4.65 \times 10^{10}$ & 3.8x \\
ethylene carbonate (1000 eV) & 47530.0 & AA & 239688.9 & 11167.3 & AA & -433.9 & -2508.4 & 9,573 & 10,646 & $4.47 \times 10^{11}$ & $1.17 \times 10^{11}$ & 3.8x \\
LiPF6 (15 eV) & 12386.9 & AA & 73258.8 & 3143.7 & AA & -380.7 & -1002.0 & 6,881 & 7,427 & $8.37 \times 10^{10}$ & $2.29 \times 10^{10}$ & 3.7x \\
LiPF6 (60 eV) & 25837.0 & AA & 146517.6 & 5831.8 & AA & -639.6 & -1656.6 & 8,537 & 9,255 & $2.17 \times 10^{11}$ & $5.30 \times 10^{10}$ & 4.1x \\
LiPF6 (250 eV) & 53575.7 & AA & 293035.2 & 12258.2 & AA & -815.4 & -2995.0 & 10,259 & 11,161 & $5.40 \times 10^{11}$ & $1.34 \times 10^{11}$ & 4.0x \\
LiPF6 (1000 eV) & 112110.7 & AA & 586070.4 & 25777.3 & AA & -1011.2 & -5663.1 & 12,047 & 13,145 & $1.33 \times 10^{12}$ & $3.33 \times 10^{11}$ & 4.0x \\
Li 2-atom (800 eV) & 1116.4 & AA & 3771.3 & 384.6 & AA & -7.7 & -124.5 & 2,849 & 3,362 & $3.12 \times 10^{9\phantom{0}}$ & $1.27 \times 10^{9\phantom{0}}$ & 2.5x \\
Li 2-atom (3000 eV) & 3356.8 & AA & 7542.7 & 1376.2 & No AA & -8.7 & -268.8 & 3,696 & 2,991 & $1.22 \times 10^{10}$ & $4.04 \times 10^{9\phantom{0}}$ & 3.0x \\
Li 2-atom (12000 eV) & 11085.2 & No AA & 15085.3 & 2823.7 & No AA & -9.0 & -557.4 & 2,860 & 3,741 & $3.11 \times 10^{10}$ & $1.04 \times 10^{10}$ & 3.0x \\
Li 16-atom (800 eV) & 44432.0 & AA & 232928.8 & 7585.8 & AA & -64.2 & -1068.7 & 7,033 & 7,778 & $3.07 \times 10^{11}$ & $5.79 \times 10^{10}$ & 5.3x \\
Li 16-atom (3100 eV) & 98922.1 & AA & 465857.5 & 16394.2 & AA & -72.3 & -2223.1 & 8,515 & 9,480 & $8.27 \times 10^{11}$ & $1.53 \times 10^{11}$ & 5.4x \\
Li 16-atom (13000 eV) & 233884.3 & AA & 931715.8 & 35719.3 & AA & -74.2 & -4532.3 & 10,033 & 11,208 & $2.30 \times 10^{12}$ & $3.93 \times 10^{11}$ & 5.9x \\
Li 54-atom (350 eV) & 312518.1 & AA & 1762362.4 & 30161.1 & AA & -185.7 & -2390.2 & 14,812 & 15,525 & $4.54 \times 10^{12}$ & $4.60 \times 10^{11}$ & 9.9x \\
Li 54-atom (1400 eV) & 648695.6 & AA & 3524724.7 & 62806.6 & AA & -237.5 & -4865.7 & 17,680 & 18,595 & $1.13 \times 10^{13}$ & $1.15 \times 10^{12}$ & 9.8x \\
Li 54-atom (5700 eV) & 1360033.9 & AA & 7049449.4 & 132571.5 & AA & -253.5 & -10182.9 & 20,614 & 21,743 & $2.75 \times 10^{13}$ & $2.83 \times 10^{12}$ & 9.7x \\
hydrogen target & 1681723.5 & AA & 9097202.2 & 103170.4 & AA & 24.1 & -4785.9 & 22,009 & 22,944 & $3.63 \times 10^{13}$ & $2.32 \times 10^{12}$ & 15.6x \\
deuterium target & 88392812.9 & AA & 496294604.4 & 2008341.5 & AA & 2685.6 & -31670.3 & 139,302 & 138,880 & $1.21 \times 10^{16}$ & $2.74 \times 10^{14}$ & 44.1x \\
carbon target & 3704637.2 & AA & 20912489.3 & 307935.3 & AA & -683.6 & -20066.4 & 30,025 & 30,622 & $1.09 \times 10^{14}$ & $9.26 \times 10^{12}$ & 11.8x \\
\hline\hline
\end{tabular}
}
\end{table}
Beyond exact gate costs, we can visualize the asymptotic speedup achieved by SOSSA by examining the effective block encoding $1$-norm for the kinetic energy and potential terms individually. We look at each term individually because the asymptotic speedup appears in different terms, $\eta$ versus $\Delta$.  For the kinetic energy term we isolate the system size independent scaling on $\Delta^{-1}$ by dividing the effective LCU 1-norm by particle count
\begin{align}
    \frac{1}{\eta}\sqrt{\Lambda_{T} (E_{\text{HF}} + \beta)}
\end{align}
and plot against $\Delta^{-1}$ on a log-log plot. For the potential terms, we isolate the effective $\eta$ scaling by multiplying by system-dependent grid spacing $\Delta$
\begin{align}
\Delta \sqrt{\left(\Lambda_{V}+ \Lambda_{X}\right)\left(E_{\text{HF}} + \beta\right)}
\end{align}
and plotting the result against $\eta$ on a log-log plot. Both plots are shown in Fig.~\ref{fig:lambda_T_VX_scaling} and compare to the LCU block encoding 1-norm scaling. Minor deviations due to the effectiveness of the Hartree-Fock upper bound for each system lead to a slight scatter in the SOSSA scalings, but the reduced asymptotic scaling in $\lambda_{\text{eff}}$ is easy to observe and recovers Eq.~\eqref{eq:sossa_lambda_effective}.

In Table~\ref{tab:ueg_qpe_cost_improvement} we show the speedup for several UEG systems compared to the LCU method of Ref.~\cite{su2021fault}. The speedup is significant and, as we noted above, it should only improve with larger system sizes. For the UEG in particular, this is important as the most useful simulations must be done at system sizes approaching the thermodynamic limit.  The UEG costs associated with the LCU algorithm of Ref.~\cite{su2021fault} are taken from Ref.~\cite{georges2025quantum} without recompiling the algorithm to take advantage of the lack of nuclear terms. It is likely if this accounting is correctly taken into account there would be a minor improvement in the LCU method in subdominant costs. 
\begin{table}[H]
\centering
\caption{Comparison of $\lambda$, Walk Operator Cost, Proxy Cost (the product of $\lambda$ and the Walk Cost), and Total Toffoli Cost with $\epsilon = \eta \times 1.6 \text{ mHa}$ for the UEG at $r_s = 5$.}
\label{tab:ueg_qpe_cost_improvement}
\begin{tabular}{llrrrrc}
\toprule
\textbf{System} & \textbf{Method} & $\lambda~~~$ & \textbf{Walk Cost} & \textbf{Proxy Cost} & \textbf{Total Cost} & \textbf{Improvement} \\
\midrule
\multirow{3}{*}{UEG-14} 
 & Ref.~\cite{georges2025quantum} (min $T$) & 153.0 & 1,230 & 188,190 & $1.32 \times 10^7$ & 1.72$\times$ \\
 & Ref.~\cite{su2021fault} & 171.0 & 1,890 & 323,190 & $2.27 \times 10^7$ & 1.00$\times$ \\
 & \textbf{SOSSA AA} & \textbf{47.1} & \textbf{1,858} & \textbf{87,432} & $6.13 \times 10^6$ & \textbf{3.70$\times$} \\
\midrule
\multirow{3}{*}{UEG-54} 
 & Ref.~\cite{georges2025quantum} (min $T$) & 4820.0 & 7,100 & 34,222,000 & $6.22 \times 10^8$ & 0.47$\times$ \\
 & Ref.~\cite{su2021fault}  & 3420.0 & 4,730 & 16,176,600 & $2.94 \times 10^8$ & 1.00$\times$ \\
 & \textbf{SOSSA AA} & \textbf{471.8} & \textbf{4,603} & \textbf{2,171,663} & $3.95 \times 10^7$ & \textbf{7.45$\times$} \\
\midrule
\multirow{3}{*}{UEG-114} 
 & Ref.~\cite{georges2025quantum} (min $T$) & 16400.0 & 9,420 & 154,488,000 & $1.33 \times 10^9$ & 0.59$\times$ \\
 & Ref.~\cite{su2021fault} & 11500.0 & 7,900 & 90,850,000 & $7.83 \times 10^8$ & 1.00$\times$ \\
 & \textbf{SOSSA AA} & \textbf{1103.0} & \textbf{7,802} & \textbf{8,605,922} & $7.41 \times 10^7$ & \textbf{10.56$\times$} \\
\bottomrule
\end{tabular}
\end{table}

\begin{figure}[H]
    \centering
    \includegraphics[width=0.5\linewidth]{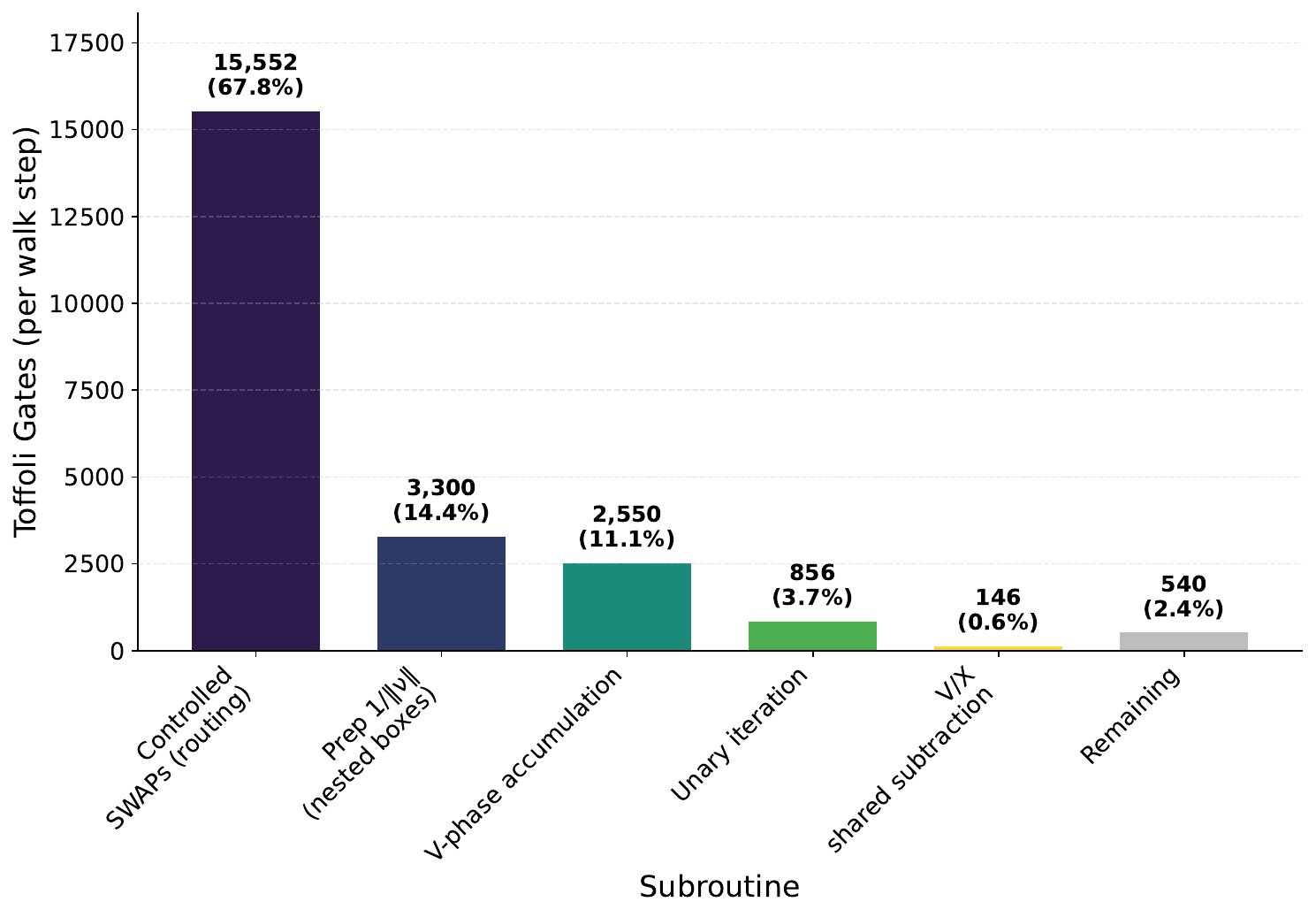}
    \caption{The top five subroutines and summed remainder costs for the Hydrogen system in Table~\ref{tab:comparison}. The $1/\|\nu\|$ preparation cost is three times the base nested-boxes cost because we use amplitude amplification for the state preparation for this system. The V-phase accumulation is the efficient Add-and-phase circuit with $\mathcal{O}(n_{p}^{2})$ described in Appendix~\ref{app:potential_compilation}.}
    \label{fig:subroutinecosts}
\end{figure}

As shown in Tables~\ref{tab:prep_costs} and~\ref{tab:select_costs}, the dominant cost is from swapping data from the electron registers into the workspace registers. This is because the number of electrons is usually significantly larger than the number of bits in the representation and the cost of swapping goes as their product while other block encoding costs are polynomial in representation size. To provide more detail, we isolate the costs of the top five subroutines to construct the walk operator for the Hydrogen system (row 18 of Table~\ref{tab:comparison}) in Figure~\ref{fig:subroutinecosts}.  Data movement accounts for an order of magnitude higher Toffoli cost than the second highest subroutine which is nested boxes state preparation. The cost given for preparation of the $1/\|\nu\|$ state is for the case where amplitude amplification is performed. 

\section{Conclusion}\label{sec:remarks}
In this work, we have constructed an analytical SOS representation of the first-quantized electronic structure Hamiltonian and shown that spectral amplification with this SOS provides both an asymptotic improvement and a practical speedup on several benchmark systems. The intuition behind the SOS form used here is that, in the plane-wave representation, the Coulomb interactions can be written as an SOS in terms of the total charge density operator. We expect this to be a low-frustration operator.
This same intuition may also be generalizable to other Hamiltonians, thereby suggesting a path to analytical SOS representations for other problems. One key advantage of this representation is that the operator can be block-encoded in a manner very similar to that of Ref.~\cite{su2021fault}. This means that the block encoding cost is comparable to the LCU method, and the reduction in the effective subnormalization constant from spectral amplification can be fully realized as a practical speedup. 

Detailed logical resource estimates for several benchmark systems were shown in Sec.~\ref{sec:gate_costs}. In all cases we found a speedup relative to the LCU method, ranging from $2 - 44\times$. It is likely that further improvements are possible from either an alternative SOS representation (see the discussion in Appendix~\ref{app:alt_sos_reps}) or from improvements to the algorithm. We found that the Toffoli cost of the algorithm is dominated by the cost of swapping data from the electron registers into the workspace. This is consistent with many other first-quantized algorithms and has a high impact on the cost. In particular, optimizing the most expensive subroutines at the level of error correction operations for a particular model architecture is necessary to get realistic physical costs. 

Applying SOSSA to closely-related problems that also use a first-quantized representation is likely to lead to similar speedups. In particular, block-encoding-based methods for electron dynamics, electron-nuclear dynamics, and ground state energy estimation with pseudopotentials can all benefit from SOS representations like the one presented in this work. Since most {\it ab initio} calculations on materials will require pseudopotentials, extending the SOSSA framework to include non-local pseudopotentials is a necessary next step.

\appendix
\section{The scaling of $\beta_{\mathrm{nuc}}$}\label{app:beta_scaling}
In this section we demonstrate that $\beta_{\mathrm{nuc}}$ in Eq.~\eqref{eq:beta-nuc} scales as $\mathcal{O}(\eta\Delta^{-1})$ for physically-relevant systems. A naive bound using the triangle inequality,
\begin{equation}
    \beta_{\text{nuc}} = \sum_{\nu \in G_0}C_\nu
     \left|\sum_{\ell}\zeta_{\ell} e^{-ik_{\nu}\cdot R_{\ell}} \right|^2 \leq \sum_{\nu \in G_0}C_\nu
     \left(\sum_{\ell}\left|\zeta_{\ell} e^{-ik_{\nu}\cdot R_{\ell}} \right|\right)^2,
\end{equation}
would suggest $\beta_{\mathrm{nuc}} \sim \eta^2$, but such a bound makes no assumptions about how ions are distributed in space. In the worst case one could imagine a system where all the ions are put into a small region of constant volume as the system is made larger and our intuition is that $\beta_{\mathrm{nuc}}$ would scale with $\eta^2$ in this case, in agreement with the bound from the triangle inequality. However, in physical systems the ions are usually distributed much more uniformly in the unit cell, leading to $\beta_{\mathrm{nuc}} \sim \eta$.

As explained in the main text, we can expand $\beta_{\text{nuc}}$ as
\begin{equation}\label{eq:beta-nuc-expanded2}
    \beta_{\text{nuc}} = \sum_{\nu \in G_0}C_\nu
     \sum_{\ell}\zeta_{\ell}^2 +
    \sum_{\nu \in G_0}C_\nu
     \sum_{\substack{\ell,\ell' \\ \ell\ne \ell'}}\zeta_{\ell}\zeta_{\ell'} e^{-ik_{\nu}\cdot (R_{\ell}-R_{\ell'})} .
\end{equation}
The first term is linear in the number of nuclei, whereas the second term is the Coulomb potential between nuclei with a background charge due to omitting $\nu=0$ in $G_0$.
We may rearrange the double sum as
\begin{equation}
    \sum_{\nu \in G_0}C_\nu
     \sum_{\substack{\ell,\ell' \\ \ell\ne \ell'}}\zeta_{\ell}\zeta_{\ell'} e^{-ik_{\nu}\cdot (R_{\ell}-R_{\ell'})} = \sum_{\substack{\ell,\ell' \\ \ell\ne \ell'}} \zeta_{\ell} \zeta_{\ell'} v({R}_{\ell} - {R}_{\ell'})\, ,
\end{equation}
where we define
\begin{equation}
    v(R) \equiv \sum_{\nu \in G_0} C_\nu e^{-i {k}_{\nu} \cdot R}  .
\end{equation}

We define a screened potential $v_\alpha(R) = \frac{e^{-\alpha \|R\|}}{2\|R\|}$, whose Fourier transform is strictly bounded at ${k}={0}$:
\begin{equation}
    \tilde{v}_\alpha({k}) = \int_{\mathbb{R}^3} \frac{e^{-\alpha \|R\|}}{2\|R\|} e^{-i {k} \cdot R} d^3R = \frac{2\pi}{\|{k}\|^2 + \alpha^2}.
\end{equation}
We can alternatively write $v(R)$ as, with the approximation that the sum over $\nu$ is taken to infinity (the continuum limit)
\begin{equation}
    v(R) \approx \lim_{\alpha \to 0} \frac{1}{\Omega} \sum_{\nu \neq \mathbf{0}} \tilde{v}_\alpha({k}_\nu) e^{i {k}_\nu \cdot R}.
\end{equation}
If $\nu=0$ were included in the sum, we would have
\begin{equation}
    \frac{1}{\Omega} \sum_{\nu \in \mathbb{Z}^3} \tilde{v}_\alpha({k}_\nu) e^{i {k}_\nu \cdot R} = \sum_{\mathbf{T}} v_\alpha(R + \mathbf{T}) .
\end{equation}
The sum without $\nu=0$ can therefore be given as
\begin{align}
    \frac{1}{\Omega} \sum_{\nu \neq {0}} \tilde{v}_\alpha({k}_\nu) e^{i {k}_\nu \cdot R} 
    &= \sum_{\mathbf{T}} v_\alpha(R + \mathbf{T}) - \frac{1}{\Omega} \tilde{v}_\alpha({0}) e^0 \nn
    &= \sum_{\mathbf{T}} \frac{e^{-\alpha \|R + \mathbf{T}\|}}{2\|R + \mathbf{T}\|} - \frac{1}{\Omega} \frac{2\pi}{\alpha^2} .
\end{align}
The subtracted constant may be rewritten as
\begin{align}
    \frac{1}{\Omega} \tilde{v}_\alpha({0}) &= \int_{\mathbb{R}^3} \frac{1}{\Omega} v_\alpha({r}) \, d^3r \nn
    &= \int_{\mathbb{R}^3} \frac{e^{-\alpha \|{r}\|}}{2 \Omega \|{r}\|} \, d^3r \nn
    &= \int_{\mathbb{R}^3} \frac{e^{-\alpha \|R - {r}\|}}{2 \Omega \|R - {r}\|} \, d^3r \nn
    &= \sum_{\mathbf{T}} \int_{\text{box } \mathbf{T}} \frac{e^{-\alpha \|R - {r}\|}}{2\Omega\|R - {r}\|} \, d^3r \, .
\end{align}

Thus $v({R})$ may be given as
\begin{align}
    v(R) &= \lim_{\alpha \to 0} \sum_{\mathbf{T}}\left[  \frac{e^{-\alpha \|R + \mathbf{T}\|}}{2\|R + \mathbf{T}\|} - \int_{\text{box } \mathbf{T}} \frac{e^{-\alpha \|R - {r}\|}}{2 \Omega \|R - {r}\|} \, d^3r \right] \nonumber \\
    &= \sum_{\mathbf{T}} \left( \frac{1}{2\|R + \mathbf{T}\|} - \int_{\text{box } \mathbf{T}} \frac{1}{2\Omega\|R - {r}\|} \, d^3r \right) .
\end{align}
Thus we obtain the expression for $v(R)$ given in Eq.~\eqref{eq:Veff}, and 
$\beta_{\text{nuc}}$ can be approximated as
\begin{align}
     \beta_{\text{nuc}} \approx \beta_{\text{nuc}}^{\rm non-BO} +\sum_{\ell} \zeta_\ell  V_{\ell}( R_{\ell} )\, ,
\end{align}
with an effective potential
\begin{equation}
    V_\ell (R_\ell) = \sum_{\ell'\ne \ell} \zeta_{\ell'} v(R_{\ell} - R_{\ell'}).
\end{equation}
Therefore, the potential for each charge is obtained by a sum over the other charges with a background charge such that the average charge is zero.
As the size of the region and number of charges is increased, the potential due to the charges at long range tends to be canceled by the background, so the potential for each charge approaches a constant.
This means that the overall $\beta_{\mathrm{nuc}}$ should be extensive.

As an alternative analysis, note that $\beta_{\mathrm{nuc}}$ is exactly the Coulomb energy per unit cell of a density, in Fourier space, of the form
\begin{equation}
    \rho(k_{\nu}) = \sum_{\ell} \zeta_{\ell} e^{-ik_{\nu}\cdot R_{\ell}},
\end{equation}
with a uniform background charge so that $\rho(0) = 0$. For most physical systems, this quantity should be extensive, but the extensivity depends on the positions of the ions. In what follows we provide evidence for this extensivity, or linear scaling in $\eta$, for typical systems.

{\bf Periodic crystals:} For supercells of a periodic crystal, $\beta_{\mathrm{nuc}}$ will be {\it exactly} linear in the system size. This can be easily seen by considering a supercell of $N_c$ copies of the primitive cell. Indicating the supercell quantities with a prime, we find that
\begin{equation}
    \beta_{\mathrm{nuc}}' = \sum_{k_{\nu}'}\frac{2\pi}{\Omega'} \frac{1}{\lVert k_{\nu}' \rVert^2} \left|\rho'(k'_{\nu})\right|^2.
\end{equation}
In the supercell there is exact phase cancellation so that the density is expanded only in the reciprocal lattice vectors of the primitive cell:
\begin{equation}
    \beta_{\mathrm{nuc}}' = \sum_{k_{\nu}}\frac{2\pi}{\Omega'} \frac{1}{\lVert k_{\nu} \rVert^2}\left|\rho'(k_{\nu})\right |^2.
\end{equation}
Note that
\begin{equation}
    \rho'(k_{\nu}) = \sum_{\ell'}\zeta_{\ell'}e^{-ik_{\nu}\cdot R_{\ell'}} = N_c\sum_{\ell}\zeta_{\ell}e^{-ik_{\nu}\cdot R_{\ell}} = N_c\rho(k_{\nu}),
\end{equation}
where we have used the fact that $R_{\ell'}$ differs from some $R_{\ell}$ by a primitive lattice translation vector. We can now express the primed quantity in terms of the unprimed quantity, and we find that
\begin{equation}
    \beta_{\mathrm{nuc}}' = \sum_{k_{\nu}}\frac{2\pi}{N_c\Omega} \frac{N_c^2}{\lVert k_{\nu} \rVert^2}\left|\rho(k_{\nu})\right |^2 = N_c \beta_{\mathrm{nuc}}
\end{equation}
as expected. Since $\eta \propto N_c$, this shows that, for supercells, $\beta_{\mathrm{nuc}} \propto \eta$. If we further use the fact that
\begin{equation}
    \sum_{\nu} \frac{1}{\lVert k_{\nu} \rVert^2} \propto \Delta^{-1},
\end{equation}
then we have that
\begin{equation}
    \beta_{\mathrm{nuc}} = \mathcal{O}(\eta \Delta^{-1}).
\end{equation}

{\bf Amorphous solids:} It is difficult to make truly precise and general statements about amorphous systems, but we can examine two limits where the scaling of $\beta_{\mathrm{nuc}}$ is clear. First we consider the limit of a perturbation about a supercell of $N_c$ primitive cells. In this case, it is straightforward to show that $\beta'$ for the larger system scales as 
\begin{equation}
    \beta_{\mathrm{nuc}}' = N_c\left[\beta_{\mathrm{nuc}} + \delta \beta\right] + \mathcal{O}(\delta R^2),
\end{equation}
where $\delta R$ is the perturbation of the positions about the perfectly periodic supercell and $\delta \beta$ is some intensive quantity that depends on $\delta R$ but not on $N_c$. In other words, for small perturbations about a periodic supercell $\beta$ is still extensive to first order in the perturbation.

\begin{figure}[H]
    \centering
    \includegraphics[width=0.3\linewidth]{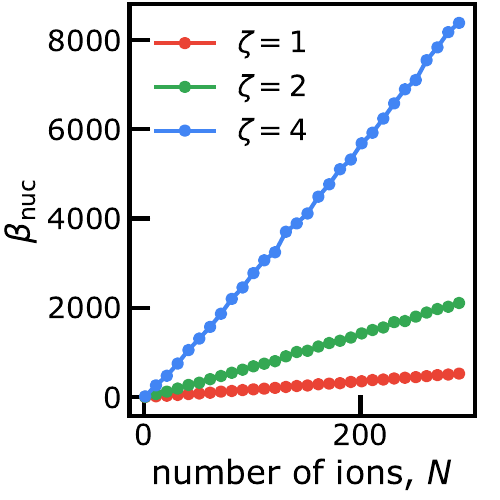}
    \includegraphics[width=0.3\linewidth]{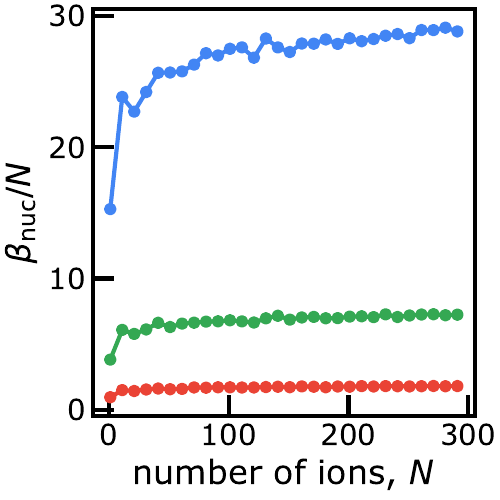}
    \caption{$\beta_{\mathrm{nuc}}$ for randomly distributed ions of charge $\zeta = 1, 2, 4$ at fixed total density and fixed kinetic energy cutoff. $\beta_{\mathrm{nuc}}$ appears to be asymptotically linear both when plotted directly (left) and when normalized by the number of ions (right). Each point is computed from an average of 100 randomly initialized configurations.}
    \label{fig:random_ions}
\end{figure}

Second, we consider the limit of randomly positioned ions. This is a reasonable approximation since, at larger length scales, the density will be indistinguishable from that of a typical amorphous system. We can easily investigate this case numerically (see Fig.~\ref{fig:random_ions}) for relevant system sizes, and we find that $\beta_{\mathrm{nuc}}\propto \eta$. We expect this to be true for any system where the local structure of the density is not changing with system size.

{\bf Molecular systems:}
For a molecular system, the unit cell contains a large vacuum region. In this case, the asymptotic behavior may not become apparent until the molecule is so large that it dwarfs the vacuum region that is needed to reduce the interaction with the periodic images. In this limit, the argument used for randomly distributed ions will apply, but such a system is likely too large for practical calculations. Precise resource estimates, like those given in Sec.~\ref{sec:logical_resources}, are a more useful metric in this case.

\section{Block Encoding}\label{app:block_encoding}
We will now write the block encoding for the full Hamiltonian. First we define the following symbols, most of which have already been defined in the main text:
\begin{itemize}
    \item $\eta$ -- number of electrons; indexed by $j$;
    \item $L$ -- number of nuclei; indexed by $\ell$;
    \item $G$ -- reciprocal lattice, $p \in G = [-\frac{N^{1/3}-1}{2}, \frac{N^{1/3}-1}{2}]^3 \subset \mathbb{Z}^3$;
    \item $N$ -- number of plane waves, $N = |G|$;
    \item $\Omega$ -- cell volume;
    \item $k_p$ -- momentum for integer vector $p$, $k_p = \frac{2\pi p}{\Omega^{1/3}}$ for $p \in G$;
    \item $\rho$ -- density, $\rho = \eta/\Omega$;
    \item $r_s$ -- Wigner-Seitz radius, $\rho^{-1} = \frac{4\pi}{3}r_s^3$;
    \item $n_p$ -- number of qubits per electron per spatial dimension, $2^{n_p} - 1 = N^{1/3}$;
    \item $w$ -- indexes the $x,y,z$ Cartesian axis direction of the plane wave;
    \item $\ket p$ -- basis states:
    \begin{align}
        \ket p &= \ket{p_x}\otimes\ket{p_y}\otimes\ket{p_z} \in \mathbb{C}^{2^{3n_p}},\\
        \ket{p_w} &= \ket{p_{w,0}}\otimes\ket{p_{w,1}}\otimes \dots\otimes \ket{p_{w,n_p-1}},\\
        p_w &= (-1)^{p_{w,n_p-1}}\sum_{r=0}^{n_p-2} 2^r p_{w,r}
    \end{align}
    \item $G_0$ -- momentum transfer $\nu \in G_0 = [-(N^{1/3}-1), N^{1/3}-1]\backslash\{(0,0,0)\} $
\end{itemize}
Note that we will be storing the wavefunction in sign-magnitude form. Later in the compilation, additions and subtractions will be performed by converting to the two's complement representation.
Recall, we need to define the kinetic energy SOS generators $T_{j,w}$ and the SOS terms to make the electron-electron self-interaction component proportional to the identity $X_{j,\nu}$. For self consistency we write out the full SOS Hamiltonian
\begin{align}
    T_{j,w} &= \frac{\sqrt{2}\pi}{\Omega^{1/3}}\sum_{p\in G} |p_w| \ket{p}\bra{p}_j\nn
    &= \frac{\sqrt{2}\pi}{\Omega^{1/3}}\sum_{p\in G}\sum_{r =0}^{n_p-2} 2^{r}  p_{w,r} \ket{p}\bra{p}_j\nn
    &= \sum_{r =0}^{n_p-2} \frac{\sqrt{2}\pi}{\Omega^{1/3}} 2^{r}  \sum_{p\in G} p_{w,r} \ket{p}\bra{p}_j\nn
    &= \sum_{r =0}^{n_p-2} \frac{\sqrt{2}\pi}{\Omega^{1/3}} 2^{r}  \sum_{p\in G} \frac{1+(-1)^{1+p_{w,r}}}{2} \ket{p}\bra{p}_j\nn
    &= \frac{1}{2}\sum_{b\in\{0,1\} }\sum_{r =0}^{n_p-2} \frac{\sqrt{2}\pi}{\Omega^{1/3}} 2^{r}  \sum_{p\in G} (-1)^{b(1+p_{w,r})} \ket{p}\bra{p}_j,\\
    U_{\text{shift},j}(\nu) &= \sum_{\substack{ p \in G: \\ p - \nu \in G}}\ket{p-\nu}\bra{p}_{j} \\
    V_{\nu} &= \sqrt{C_{\nu}}\left(\sum_{j=1}^{\eta}U_{\text{shift},j}(\nu) - \sum_{\ell=1}^{L}\zeta_{\ell}e^{-i k_{\nu}\cdot R_{\ell}}\mathbb{I}\right) = \sqrt{C_{\nu}}\tilde{V}_{\nu} \;\;,\;\;C_{\nu} = \frac{2\pi}{\Omega \|k_{\nu}\|^{2}} \nn
    &= \sqrt{C_{\nu}}\left(\sum_{j=1}^{\eta}U_{\text{shift},j}(\nu) - \sum_{m=1}^{Z_{\text{tot}}}e^{-i k_{\nu}\cdot R_{\ell}(m)}\mathbb{I}\right) \\
X_{j,\nu}  &= \sqrt{\frac{2\pi}{\Omega}} \frac 1{\|k_\nu\|} \sum_{\substack{p \in G: \\p-\nu \notin G}}  \ket{p-\nu} \bra{p}_j = \sqrt{C_{\nu}}\sum_{\substack{p \in G: \\p-\nu \notin G}}  \ket{p-\nu} \bra{p}_j \, .
\end{align}
We have expanded the summation over the nuclei in the $V_{\nu}$ to a larger index $m$ which runs over the total charge but is no longer weighted by $\zeta_{\ell}$.  The notation $R_{\ell}(m)$ means that for each index $m$ we map to a particular $R_{\ell}$ value for a nucleus. This rewriting ultimately leads to a simplification in the block encoding and allows us to overload the register for $j$ and $Z_{\text{tot}}$.

\subsection{Modification for $\nu$-state prep failure}\label{app:be_nu_state_prep}

Highlighted in Refs~\cite{su2021fault, babbush2019quantum}, the nested boxes state preparation for the outer state over the momentum difference register $\nu$ has a failure rate because it uses rejection sampling to prepare a discrete probability distribution. The failure rate is approximately $1/4$, so a single round of amplitude amplification can boost the success probability close to one at the expense of a longer circuit (approximately three times longer). Thus, the authors of Ref.~\cite{su2021fault} proposed that upon state preparation failure they block encode the kinetic energy, saving an artificial increase in the LCU 1-norm.  We utilize the same logic here for preparing the kinetic energy components of the SOS and the potential components of the SOS.
In the following we summarize the procedure adapted from that in Ref.~\cite{su2021fault}.

To implement this procedure correctly, we need to ensure that the block encodings of $T/\Lambda_T$, $V/\Lambda_V$, and $X/\Lambda_X$ are combined such that we are block encoding
\begin{equation}
    \frac{T+V+X}{\Lambda_T+\Lambda_V+\Lambda_X}.
\end{equation}
Since the block encodings of $V$ and $X$ both use the prepared $\nu$ state, for the purpose of this discussion we regard those as both being block encoded together as
\begin{equation}
    \frac{V+X}{\Lambda_V+\Lambda_X},
\end{equation}
and explain how to combine them with the appropriate weighting of $T$.

First, note that if we were to just block encode $T$ for all failure cases of the preparation, then we would be block encoding an operator proportional to
\begin{equation}
    p_\nu \frac{V+X}{\Lambda_V+\Lambda_X} + (1-p_\nu) \frac{T}{\Lambda_T},
\end{equation}
where $p_\nu$ is the probability of success of the preparation.
For this to be the appropriate weighting between $V+X$ and $T$, then we would need
\begin{equation}
    \frac{p_\nu}{\Lambda_V+\Lambda_X} = \frac{1-p_\nu}{\Lambda_T},
\end{equation}
which implies
\begin{align}\label{eq:lambda_ratio}
    \frac{(1 - p_{\nu})}{p_{\nu}} = \frac{\Lambda_{T}}{\Lambda_{V} + \Lambda_{X}}.
\end{align}
In the following we denote the qubit with the result of the success of the state preparation by $\ket{f_\nu}$.
In practice there will not be perfect equality, so we introduce an additional qubit to throttle the weighting of either the $T$ branch or the $V+X$ branch. 
In order to appropriately set the rotation for the ancilla qubit we have two scenarios to consider:
\begin{enumerate}
    \item[1.] $\frac{\Lambda_{T}}{\Lambda_{V} + \Lambda_{X}} < \frac{(1 - p_{\nu})}{p_{\nu}}$ corresponds to a case where the failure during the $\nu$ state preparation is sufficient to cover all the weight for $T$ and then some. In other words, the circuit produces more trash to be recycled than is needed (in terms of $\Lambda_{T}$). This is handled by throttling the application of $T$ by imposing that not only should we have a failure in state preparation of the nested boxes state but we also have the ancilla qubit in state $|0\rangle$. This is denoted the AND strategy because we require the AND of the correct states in the failure flag and ancilla qubit.  
    
    For this scenario we must set the ancilla qubit $|q_{T}\rangle = \cos(\theta)|0\rangle+\sin(\theta)|1\rangle$ angles appropriately. We can determine $\theta$ by solving
    \begin{align}
    \frac{(1-p_{\nu})\cos^{2}(\theta)}{p_{\nu}} = \frac{\Lambda_{T}}{\Lambda_{V} + \Lambda_{X}}
    \end{align}
    where the numerator is a product of the probability for $|f_{\nu}\rangle = |1\rangle$ and $|q_{T}\rangle=|0\rangle$. Note that the condition $p_\nu\Lambda_{T}<(1-p_\nu)(\Lambda_{V}+\Lambda_{X})$ implies that the choice of $\cos\theta$ is less than 1. 
    
    To determine the effective 1-norm ($\lambda_{\text{eff}}$) of the block encoding for this AND strategy, we examine the global scaling factor required to map the physical probabilities of the circuit back to the target Hamiltonian $H = T + V + X$. Because the potential branch ($V+X$) is applied unconditionally upon the success of the nested boxes, its physical weight in the LCU is exactly $p_{\nu}$. The raw block-encoded operator produced by the circuit is therefore:
    \begin{align}
        p_{\nu} \left( \frac{V+X}{\Lambda_{V} + \Lambda_{X}} \right) + (1-p_{\nu})\cos^{2}(\theta) \left( \frac{T}{\Lambda_{T}} \right).
    \end{align}
    By substituting our balanced choice of $\theta$, we can factor out the coefficient of the potential terms globally to give
    \begin{align}
        \left( \frac{p_{\nu}}{\Lambda_{V} + \Lambda_{X}} \right) \Big( V + X + T \Big) = \frac{H}{\lambda_{\text{eff}}}.
    \end{align}
    Taking the reciprocal of this global prefactor yields the effective 1-norm for the AND strategy as
    \begin{align}
        \lambda_{\text{eff}} = \frac{\Lambda_{V} + \Lambda_{X}}{p_{\nu}}.
    \end{align}
    As established in Ref.~\cite{su2021fault} Appendix D, this result demonstrates that in the AND regime, the kinetic energy operator $T$ is simulated with zero additional overhead. Because $\Lambda_{T}$ fits entirely within the `trash' amplitude generated by the $(1-p_{\nu})$ failure rate, it drops out of the effective 1-norm completely.
    \item[2.] $\frac{\Lambda_{T}}{\Lambda_{V} + \Lambda_{X}} \geq \frac{(1 - p_{\nu})}{p_{\nu}}$ corresponds to the case where failure during the $\nu$ state preparation is insufficient to cover all the weight needed for applying the $T$ operator.  In other words, there is \textit{too little trash} generated by nested-boxes state preparation to be recycled into $T$ to cover all of $T$. In this case we consider applying $T$ if nested-boxes state preparation fails OR $|q_{T}\rangle = |0\rangle$.
    
    The corresponding block-encoded operator is then
    \begin{align}
        p_{\nu}\sin^2(\theta) \left( \frac{V+X}{\Lambda_{V} + \Lambda_{X}} \right) + [\cos^2(\theta) + (1-p_{\nu})\sin^{2}(\theta)] \left( \frac{T}{\Lambda_{T}} \right).
    \end{align}
    The factor of $p_{\nu}\sin^{2}(\theta)$ is because the $V+X$ operator is applied whenever $|q_{T}\rangle=|1\rangle$ and nested boxes succeeds.
    For $T$, the factor of $\cos^{2}(\theta) + (1-p_{\nu})\sin^{2}(\theta)$ is because $T$ is applied when $|q_{T}\rangle = |0\rangle$ or the nested boxes state preparation fails.
    As a result, for the correct combination of operators we need
    \begin{equation}
         \frac{p_{\nu}\sin^2(\theta)}{\Lambda_{V} + \Lambda_{X}}  =  \frac{\cos^2(\theta) + (1-p_{\nu})\sin^{2}(\theta)}{\Lambda_{T}} .
    \end{equation}
    Rearranging gives
    \begin{align}
        \sin^{2}(\theta) = \frac{\Lambda_{V} + \Lambda_{X}}{p_{\nu}\left(\Lambda_{T} + \Lambda_{V} + \Lambda_{X}\right)}.
    \end{align}

    In this case, we either block encode $V+X$ or $T$ and not the identity, so no amplitude is wasted, and the block-encoded operator is
    \begin{align}
     \frac{T + V + X }{\Lambda_{T} + \Lambda_{V} + \Lambda_{X}}  = \frac{H}{\lambda_{\text{eff}}}.
    \end{align}
    Taking the reciprocal of this global prefactor yields the effective 1-norm for the OR strategy as
    \begin{align}
        \lambda_{\text{eff}} = \Lambda_{T} + \Lambda_{V} + \Lambda_{X}.
    \end{align}
\end{enumerate}

To integrate this failure-recycling trick into the SOSSA protocol, we must account for a structural difference. The SOSSA kinetic operator $T_{j,w}$ has an outer summation over indices $\{w, j\}$ and an inner summation $\{r,b\}$. If the $\nu$-state preparation fails and we redirect the branch to apply $T$, the $w$ and $r$ registers must already be properly prepared. There are no summations over these indices in the other SOS operators and thus we can prepare  $w$ and $r$ uncontrolled on the branch register $S$. 

Optionally, we can perform amplitude amplification to boost the probability of success for $1/\|\nu\|$ preparation.  After a single round, the probability is 
\begin{align}
    p_{\nu}^{\text{amp}} =  \sin^{2}\left(3 \sin^{-1}\left(\sqrt{p_{\nu}}\right)\right)
\end{align}
which will re-weight the case where we have the AND-protocol and result in performing the OR-protocol. The Toffoli cost for nested boxes, $3n_p^2 + 11n_p +2 + 4n_{\mathcal{M}}(n_p + 1)$, is increased by a factor of three if performing this step to minimize total gate cost.
Note that the cost of the procedure as given in Ref.~\cite{su2021fault} can be reduced slightly by a different method of performing the controlled Hadamards.
Rather than explicitly performing a controlled Hadamard, we can perform a controlled swap between a $\ket +$ state and a $\ket 0$ state.
This is because the controlled Hadamard only need be performed on a $\ket 0$ state.
As a result measurement-based uncomputation can be used for this part of the procedure as well, and there is no extra Toffoli cost in $P_{\text{out}}^\dagger$.

\subsection{State Preparation for Selection between $T, V$, and $X$}\label{app:exec_flag_calc}
The logical routing depends entirely on whether the target Hamiltonian falls into the AND or OR recycling regime.
The choice of which part of the Hamiltonian to block encode is governed by the flags $\text{Exec}_{V}$, $\text{Exec}_{X}$, and $\text{Exec}_{T}$.
The Boolean logic used to compute these flags is determined classically during compilation and implemented as follows:

\textbf{Regime 1: The AND Strategy} \\
The flags can be computed as
\begin{enumerate}
    \item $\text{Exec}_{T}$ is computed as $f_{\nu} = 1$ AND $q_{T} = 0$
    \item $\text{Exec}_{V}$ is computed as $f_{\nu} = 0$ AND $q_{VX} = 0$
    \item $\text{Exec}_{X}$ is computed as $f_{\nu} = 0$ AND $q_{VX} = 1$
\end{enumerate}
Only two Toffolis are needed, because it is possible to copy the NOT of the value of $f_{\nu}$ and use $q_{VX}$ to control the swap of the qubits storing $\text{Exec}_{V}$ and $\text{Exec}_{X}$.

\textbf{Regime 2: The OR Strategy} \\
The execution flags are computed as:
\begin{enumerate}
    \item $\text{Exec}_{T}$ is computed as $f_{\nu} = 1$ OR $q_{T} = 0$
    \item $\text{Exec}_{V}$ is computed as $\text{Exec}_{T}=0$ AND $q_{VX}=0$
    \item $\text{Exec}_{X}$ is computed as $\text{Exec}_{T}=0$ AND $q_{VX}=1$
\end{enumerate}
Again only two Toffolis are needed.
One Toffoli is used to compute the OR for $\text{Exec}_{T}$.
The NOT of that is copied into the one of the qubits for $\text{Exec}_{V}$ and $\text{Exec}_{X}$ with a CNOT, then $q_{VX}$ is used to control the swap of the 1 into the correct qubit.

In both cases, the flag qubits do not require Toffoli gates to uncompute.
The Toffolis used above are applied to qubits in the $\ket{0}$ state, so the process can be inverted using measurements and phase corrections.
Note that we retain the control qubits ($q_T, q_{VX}, f_\nu$) through the SELECT oracle.

\subsection{Structure of \textsc{PREP}$_{\text{out/in}}$}\label{app:prep_inout_be}
The outer state preparation includes the signal state preparation for the $\nu$-register, $\{q_{T}, q_{VX}\}$ control registers, execution flag preparation, and unconditional preparation over index $w$ and $r$. The $w$-bit preparation takes $3n + 2b_{r} - 9$ with $b_{r}=8$ and $n=2$ yielding $13$ Toffolis.
The superposition over $r$ may be implemented with $n_{p}-2$ controlled Hadamards applied with controlled swaps, which can be uncomputed with measurement-based uncomputation.
The cost of the preparation for $\{\nu, q_{T},q_{VX},\text{Exec}_{T/V/X}\}$ is described in the previous section and occurs in three steps: 1) the preparation of the two selection qubits ($q_{T}$ and $q_{VX}$) to $n_T$ bits of precision requiring $2n_T - 6$ Toffolis and 2) the nested boxes state preparation costing $ 3  n_p^2 + 11 n_p + 2 + 4 n_{\mathcal{M}}  (n_p + 1)$ 
Toffolis and returning the failure flag $|f_{\nu}\rangle$ followed by 3) computing the flags $\text{Exec}_{T/V/X}$ costing $2$ Toffolis for either the AND or OR strategies based on the ratio $\Lambda_{T} / (\Lambda_{V} + \Lambda_{X})$. The appropriate uncomputation costs are documented in Table~\ref{tab:prep_costs}.

The next phase of the preparation we will focus on is the remaining outer preparation over the $j$-register and the inner preparations. To avoid excessive controls and additional registers in order to implement the correct outer and inner prepare states for each SOS generator we utilize a combination of an overloaded register for sums over the electron and charge indices ($j$, $m$), controlled state preparations, and controlled reflections $(2\Pi_{\text{in}} - \mathbb{I})$ in order to block encode the appropriate terms. The only non-trivial inner preparation is for the $V_{\nu}$ term which has two LCUs involving index $j$ and index $\ell$.  Because there are two terms we prepare a flag qubit $f_\text{nuc-flag}$ to split the amplitude between the electron-electron interaction ($|0\rangle$) and the electron-nuclear interaction ($|1\rangle$). Because this split is only required by the potential operator, the rotation is controlled on $\text{Exec}_V = |1\rangle$. For the $T$ and $X$ branches, this operation is bypassed, and $f_{\text{nuc-flag}}$ naturally remains $|0\rangle$.  

For a simulation involving charged system where $\eta \neq Z_{\text{tot}}$ the sizes of the superpositions on \texttt{idx} for the $j$ and $Z_{\text{tot}}$ branches of $V_{\nu}$ differ. Thus, depending on a flag ($\text{flag-nuc}$), we prepare a different size superposition on the register \texttt{idx}. First, we review how to make uniform state preparation over $\eta$ values controlled.
The controlled state prep proceeds by Hadamards to create an equal superposition, an inequality test, rotation of an ancilla qubit, reflection about the joint result, inversion of the rotation, inequality test and Hadamards, reflection about zero, then the Hadamards and inequality test again~\cite{su2021fault, PRXQuantum.1.020312}.
Instead of having an additional register for $Z_{\text{tot}}$ we utilize the joint register \texttt{idx} requiring a joint controlled uniform superposition.  What then needs to be controlled is the following:
    \begin{enumerate}
        \item A number of Hadamards corresponding to the difference in the number of bits for the two numbers $n_{\eta}$ bits for $\eta$ and $n_{Z_{\text{tot}}}$ bits for $Z_{\text{tot}}$.
        These Hadamards need to be made controlled 3 times, though one is an inversion.
        Because the controlled Hadamards are performed on $\ket 0$ states, they may be performed with controlled swaps, and the inversion may be performed with measurement-based uncomputation, as noted in Appendix \ref{app:be_nu_state_prep}.
        The Toffoli cost is therefore $2|n_\eta-n_{Z_{\text{tot}}}|$.
        \item The inequality test needs to be with a number output into a quantum register.
        Performing the inequality test with a quantum register instead of a classically chosen number increases the cost by 1 Toffoli.
        The inequality test is performed twice to make it two extra Toffolis.  (The inversion of the inequality test is performed with Cliffords.)
        \item The rotation on the ancilla qubit also needs to be by an angle stored in a quantum register, increasing the Toffoli cost by 1.
        The rotation is performed twice, increasing the Toffoli count by 2.
    \end{enumerate}
Overall, the Toffoli cost to control between preparing an equal superposition of two different numbers is increased by $2|n_\eta-n_{Z_{\text{tot}}}|+4$ resulting in a total cost for the uniform state preparations of:
\begin{align}
3n_\eta + 2b_r - 9 + 2|n_\eta-n_{Z_{\text{tot}}}|+4.
\end{align}
Just as in the neutral case we would need to call this preparation four times for the total walk operator.

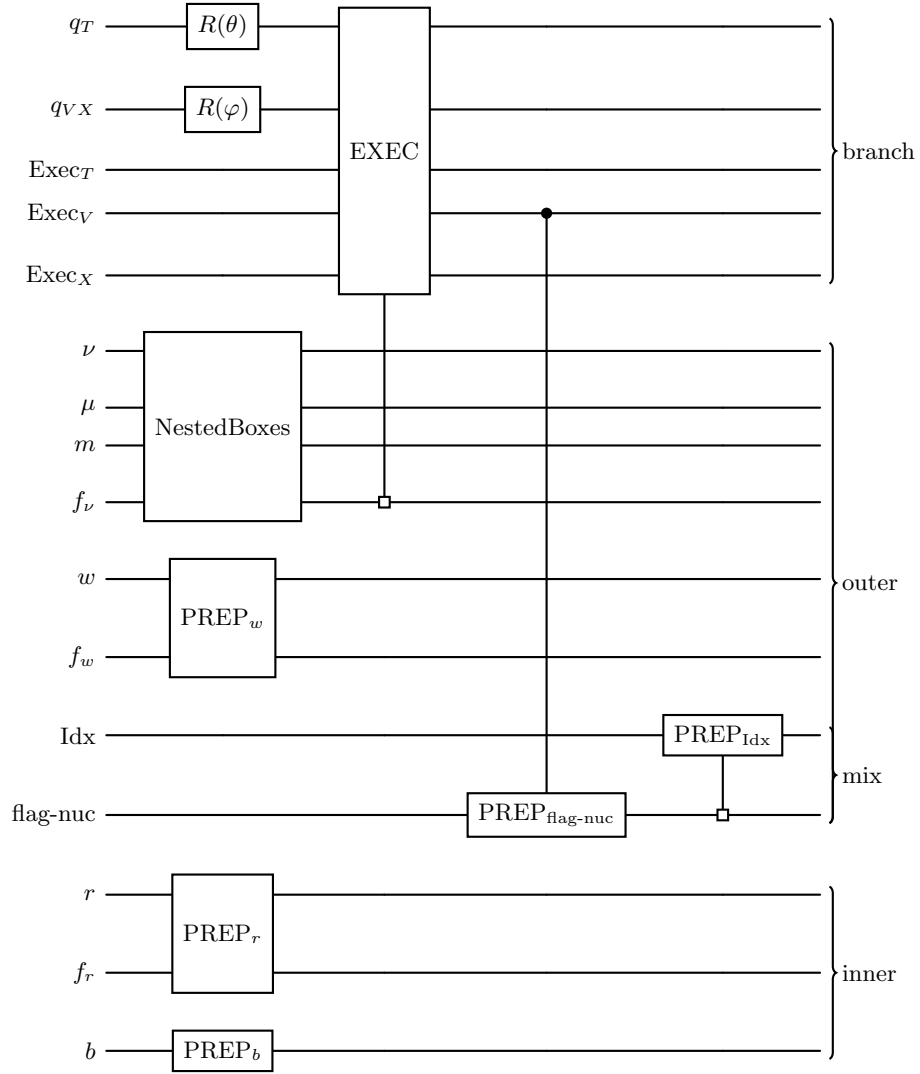
\begin{figure}
    \centering
    \begin{quantikz}
    \lstick{$q_T$}
        & \gate{R(\theta)}
        & \gate[5]{\text{EXEC}} \vqw{8}
        &
        &
        &
        \rstick[wires=5]{branch}\\
    \lstick{$q_{VX}$}
        & \gate{R(\varphi)}
        & 
        &
        &
        &
        \\
    \lstick{Exec$_T$}
        & 
        &
        & 
        &
        &
        \\
     \lstick{Exec$_V$}
        & 
        &
        & \control{}
        &
        &
        \\
     \lstick{Exec$_X$}
        & 
        &
        &
        &
        &
        \\
    \lstick{$\nu$}
        & \gate[4]{\text{NestedBoxes}}
        &
        & 
        & 
        & 
        \rstick[wires=8]{outer}\\
    \lstick{$\mu$}
        & 
        & 
        &
        &
        &
        \\
    \lstick{$m$}
        & 
        & 
        &
        &
        &
        \\
    \lstick{$f_{\nu}$}
        &
        & |[operator]| 
        &
        &
        &
        \\
     \lstick{$w$}
        & \gate[2]{\text{PREP}_w}
        & 
        &
        &
        &
        \\
    \lstick{$f_{w}$}
        &
        & 
        &
        &
        &
        \\
    \lstick{Idx}
        &
        & 
        &
        & \gate{\text{PREP}_{\text{Idx}}}\vqw{1}
        &
        \rstick[wires=2]{mix}\\
    \lstick{flag-nuc}
        &
        & 
        & \gate{\text{PREP}_{\text{flag-nuc}}}\vqw{-9}
        &|[operator]| 
        &
        \\
    \lstick{$r$}
        &\gate[2]{\text{PREP}_r}
        & 
        &
        &
        &
        \rstick[wires=3]{inner}\\
    \lstick{$f_{r}$}
        &
        & 
        &
        &
        &
        \\
    \lstick{$b$}
        & \gate{\text{PREP}_b}
        & 
        &
        &
        &
        \\
    \end{quantikz}
    \caption{Structure of the mixed outer/inner prepare $P_{\text{out/in}}$. The labels on the right-hand-side are for comparison with Fig.~\ref{fig:circuit}.}
    \label{fig:circuit_P_oi}
\end{figure}

\begin{figure}
    \centering
    \begin{quantikz}
    \lstick{Exec$_T$}
        & 
        &
        &
        &
        &
        \\
     \lstick{Exec$_V$}
        & 
        &\control{}
        &
        &
        &
        \\
     \lstick{Exec$_X$}
        & 
        & 
        &
        &
        &
        \\
    \lstick{Idx}
        &
        & 
        & \gate{\text{PREP}_{\text{Idx}}}\vqw{1}
        &
        &
        \rstick[wires=2]{mix}\\
    \lstick{flag-nuc}
        &
        & \gate{\text{PREP}_{\text{flag-nuc}}}\vqw{-3}
        &|[operator]| 
        &
        &
        \\
    \lstick{$r$}
        &\gate[2]{\text{PREP}_r}
        & 
        &
        &
        &
        \rstick[wires=3]{inner}\\
    \lstick{$f_{r}$}
        &
        & 
        &
        &
        &
        \\
    \lstick{$b$}
        & \gate{\text{PREP}_b}
        & 
        &
        &
        &
        \\
    \end{quantikz}
    \caption{Structure of inner prepare $P_{\text{in}}$. Labels on the right-hand-side are for comparison with Fig.~\ref{fig:circuit}.}
    \label{fig:circuit_P_i}
\end{figure}
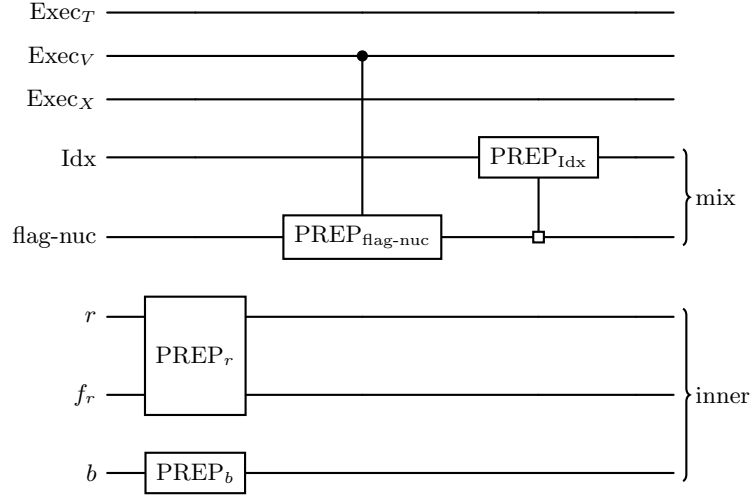

In Table~\ref{tab:prep_costs}, we provide a comprehensive breakdown of the Toffoli gate costs for every component of the merged outer and inner state preparation ($P_{\text{out/in}}$). The table explicitly separates the forward generation from the uncomputation for a single call, and provides the number of times each subroutine is queried by the SOSSA walk operator. The structure of the merged (out/in) state preparation circuit is shown in Fig.~\ref{fig:circuit_P_oi} and the circuit for just the inner preparation is shown in Fig.~\ref{fig:circuit_P_i}.

\subsection{Select implementations}
\subsubsection{Kinetic Energy}\label{app:kinetic_energy}
This is very similar to Ref.~\cite{su2021fault} in that we expand the absolute value in a sum over the binary representation of the magnitude part of the signed integer to construct the LCU
\begin{align}
T_{j,w} = \frac{1}{2} \sum_{b \in \{0,1\}} \sum_{r=0}^{n_p-2} \frac{\sqrt{2}\pi}{\Omega^{1/3}} 2^r \sum_{p \in G} (-1)^{b(1+p_{w,r})} |p\rangle \langle p|_j.
\end{align}
The difference from Ref.~\cite{su2021fault} is that we only consider one bit from $p$, rather than a product of bits.
This is because the square is obtained from the oblivious amplitude amplification, rather than being performed using bit products as in Ref.~\cite{su2021fault}.
Implementation of this LCU has the following structure:
\begin{enumerate}
   \item From the $3n_p$-bit workspace register, controlled by the spatial index $|w\rangle \in \{0, 1, 2\}$, we copy the corresponding $n_{p}-1$ magnitude bits into a temporary 1D register.
   In Ref.~\cite{su2021fault} its cost was given as $3(n_{p}-1)$ Toffoli gates, but it may be reduced to $2(n_{p}-1)$.
   The reason is that we can copy the first component without the control, and XOR it into the other two components with CNOTs.
   Then we control the copy of the other two components into the output register with cost $n_{p}-1$ each.
   Note that $w$ may be assumed to be given in one-hot unary because the conversion between binary and unary can be performed with Cliffords in this case.
   \item From the temporary register, controlled by the bit-index $|r\rangle$, we copy the $r$-th bit $p_{w,r}$ into a dedicated ancilla qubit $q_{r}$.
   In Ref.~\cite{su2021fault} the cost was given as $n_{p}-1$, but it may be reduced to $n_{p}-2$ using the same method as before (copy the first without control).
   \item Finally, we must apply the LCU phase $(-1)^{b(1+p_{w,r})}$. The $|b\rangle$ register is prepared in the $|+\rangle_{b}$ state. This phase dictates that a global $-1$ phase should be applied if and only if $b=1$ and $p_{w,r}=0$. We achieve this by applying an $X$ gate to $q_r$, followed by a CCZ gate controlled by $|b\rangle$, $q_r$, and the execution flag $\text{Exec}_{T} = |1\rangle$, and a final $X$ gate to restore $q_r$. This requires $1$ Toffoli gate for the CCZ.
   \item The ancilla qubits may be erased with measurements and Clifford gates, as in Ref.~\cite{su2021fault}
\end{enumerate}
\textit{Total Cost:} The kinetic arithmetic requires $3(n_p - 1)$ Toffolis.
\\
\subsubsection{Potential $V$}\label{app:potential_compilation}
The potential term has two components that are activated based on the value of $f_{\text{nuc-flag}}$: 1) a momentum shift and 2) phasing. 
\begin{enumerate}
    \item If $f_{\text{nuc-flag}}=0$ we apply the momentum shift for the electron branch. This physically implements $U_{\text{shift}, j}(\nu)$. Note we will use the overflow bits to flag $(p - \nu) \notin G$.
    \begin{enumerate}
        \item Convert $q_{w}$ to two's complement, which costs $n_{p}-2$ Toffolis.
        \item Perform a controlled swap of $\nu_{w}$ (controlled by $\text{Exec}_{X} \vee (\text{Exec}_{V} \wedge \neg f_{\text{nuc-flag}})$) into a temporary ancilla $|\tilde{\nu}\rangle$ costing $n_{p}+1$ Toffoli. Uncomputing uses measurement-based uncomputation because we assume the $\tilde{\nu}_{w}$ register is originally $|0\rangle$, and for this input CWAP is decomposed into AND followed by a CNOT. Preparing the control qubit costs $1$ additional Toffoli.
        \item Do the in-place addition/subtraction (subtraction handled by CNOTS and the carry-in wire) costing $n_{p}+1$ Toffolis.
        \item Convert $q_{w}$ back to sign-magnitude encoding, costing $n_{p}$ Toffolis.
    \end{enumerate}
    The total cost is thus $4 n_{p}+1$ Toffolis for a single direction.
    The $+1$ is from preparing the control qubit in part (b) which need only need be done once, so the total cost for all 3 directions is $12 n_{p} + 1$ Toffolis. 
    \item If $f_{\text{nuc-flag}}=1$:
    As each component of $\nu$ is swapped in 1(b) above, we also apply the phasing for the corresponding component for the $f_{\text{nuc-flag}}=1$ branch.
    This is because the control of the phase is implemented by this controlled swap.
    The value of $\tilde{R}_{\ell,w}$ is output from the iteration over $\max[\eta, Z_{\text{tot}}]$, so does not require additional Toffolis.
        As in Ref.~\cite{su2021fault}, We use $\tilde{R}_{\ell} = 2\pi R_{\ell}/\Omega^{1/3}$ instead of $R_{\ell}$ to enable the efficient multiply-accumulate circuit to compute $k_{\nu,w} R_{\ell,w}$ which is directly added into the phase gradient register.
        After doing this for each component, this applies $e^{-i k_\nu \cdot R_\ell}$ to the state. The leading order cost is $6 n_{p} n_{R}$ Toffoli with a catalytic phase gradient state~\cite{su2021fault}. A detailed accounting of the multiply-accumulate circuit is included in Ref.~\cite{su2021fault} Section IID. The full expression is provided in Eq.~(97) of Ref.~\cite{su2021fault} giving the following Toffoli costs
        \begin{align}
            \begin{cases}
                3\left[2 n_{p}n_{R} - n_{p}(n_{p}+1) - 1 \right], & n_{R} > n_{p} \\
                3n_{R}\left(n_{R} - 1 \right), & n_{R} \leq n_{p}.
            \end{cases}
        \end{align}
        \item Apply a $Z$ gate to $|f_{\text{nuc-flag}}\rangle$ to kick back the $-1$ phase.
    This physically implements $-U_{\text{phase}, \ell}(\nu)$.
\end{enumerate}

The SOSSA walk operator requires executing the block encoding unitary twice per step ($U_R$ forward, and $U_R^\dagger$ in reverse). This means $S_{\text{in}}$ must apply the phase $e^{-i k_\nu \cdot R_\ell}$, while $S_{\text{in}}^\dagger$ must apply the conjugate phase $e^{+i k_\nu \cdot R_\ell}$. 

\begin{figure}
    \centering
    \begin{quantikz}
    \lstick{Exec$_T$}
        &
        &\control{}
        &
        &
        &
        &
        &
    \\
    \lstick{Exec$_V$}
        &
        &
        & \gate[2]{\text{OR}}
        & \control{}
        &
        &
        &
    \\
    \lstick{Exec$_X$}
        &
        &
        &
        &
        &\control{}
        &
        &
    \\
    \lstick{$\nu$}
        &
        &
        &|[operator]|
        &|[operator]|
        &
        &
        &
    \\
    \lstick{$w$}
        & 
        &|[operator]|
        &
        &
        &
        &
        &
    \\
    \lstick{Idx}
        &|[operator]| 
        &
        &
        &
        &
        &|[operator]| 
        &
    \\
    \lstick{flag-nuc}
        &
        &
        &|[operator]|
        &|[operator]|
        &
        &
        &
    \\
    \lstick{$r$}
        &
        &|[operator]|
        &
        &
        &
        &
        &
    \\
    \lstick{$b$}
        &
        & \control{}
        &
        &
        &
        &
        &
    \\
    \lstick{overflow}
        &
        &
        & \gate[2]{\text{subtraction}}\vqw{-7}
        &
        & \gate{\text{X-check}}\vqw{-7}
        &
        &
    \\
    \lstick{workspace}
        & \gate[3]{\text{Routing}}\vqw{-5}
        & \gate{\text{kinetic phase}}\vqw{-10}
        &
        & \gate[2]{\text{V-phase}}\vqw{-9}
        &
        &\gate[3]{\text{Routing}^{\dagger}}\vqw{-5}
        &
    \\
    \lstick{$R_{\ell}$}
        &
        &
        &
        &
        &
        &
        &
    \\
    \lstick{$\ket{\Psi}$}
        &
        &
        &
        &
        &
        &
        &
    \\
    \end{quantikz}
    \caption{The structure of the select oracle ($S$). The `kinetic phase' block is described in Appendix~\ref{app:kinetic_energy}. The `subtraction', `V-phase', and `X-check' blocks implement the potential terms as described in Appendix~\ref{app:potential_compilation}.}
    \label{fig:circuit_select}
\end{figure}
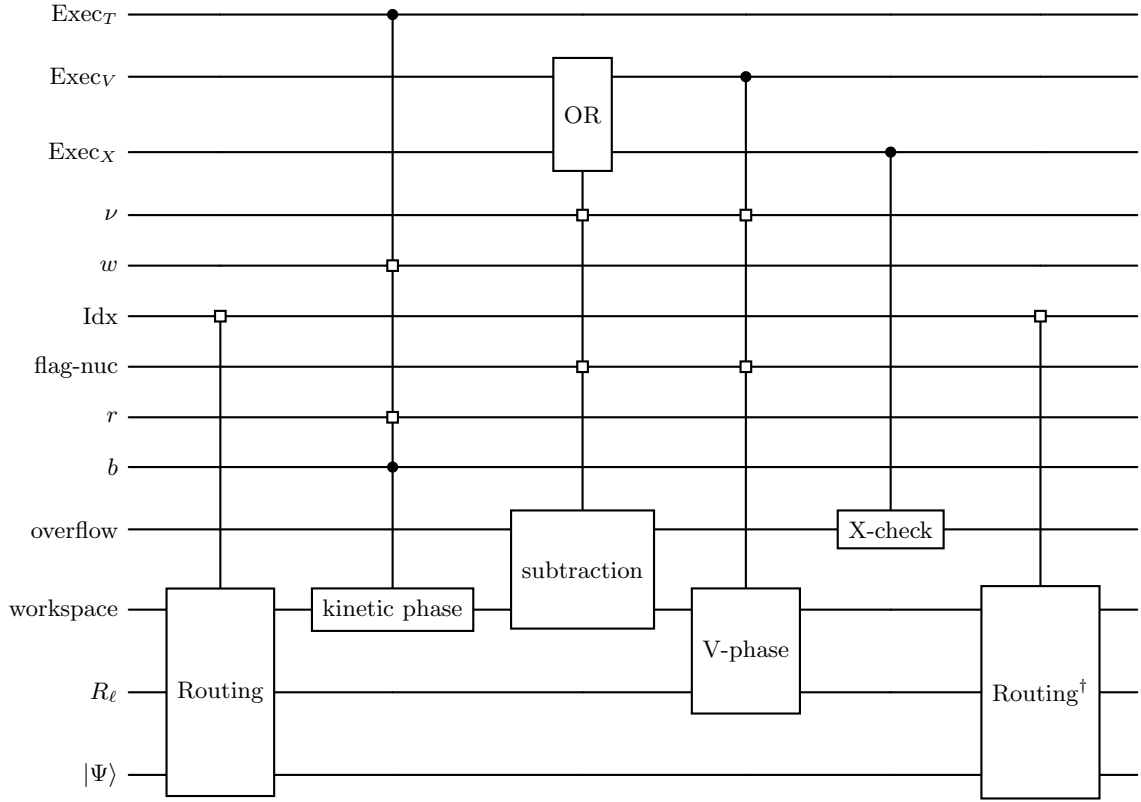

Because the overflow bits flag $(p-\nu)\notin G$, we need to reflect on these bits being zero.
Due to overflow logic, the two overflow bits in each direction are never simultaneously in the $|1\rangle$ state, so we can reduce the number of bits we need to reflect on.
In each direction, perform a CNOT between the two overflow bits to give a single bit flagging overflow, then apply a multiply-controlled Toffoli controlled by those 3 bits to flip an additional ancilla flag.
It is this ancilla flag that is reflected upon during the inner qubitization step. The cost is two additional Toffolis.
This can be erased at zero cost with measurement-based uncomputation.
The overall logic of the select oracle is shown in Fig.~\ref{fig:circuit_select}, though without showing our method of implementing the control by swapping $\nu$ with $\tilde\nu$.

\subsection{Reflection costs}\label{app:ref-section}
Recall that, because of our flag qubits, we have a direct-sum definition of the inner Hilbert space
\begin{align}
    \Pi_{\text{in}} &= \sum_{s=0}^{2} |s\rangle\langle s|_{S} \otimes \mathbb{I}_{\text{out}}^{(s)} \otimes |0\rangle\langle 0|_{\text{in}}^{(s)} .
\end{align}
The inner Hilbert space qubits $\mathcal{H}_{\text{in}}^{(s)}$ must be reflected upon, and thus we must use multi-controlled $Z$ gates controlled by the execution flags. 
The execution bits, which are one-hot encoded $\text{Exec}_{T,V,X}$, are in the outer preparation and can be used to control the reflection application. Thus we just need to count how many bits we need to reflect on for each flag. 

We now recount every single qubit that defines the state $P_{\text{out/in}} |0\dots0\rangle$, so we know exactly how large the $\Pi_{\text{all}}$ multi-controlled $Z$ gate is.

\begin{enumerate}
    \item[1.] Branch \& Logic Qubits:
    \begin{enumerate}
        \item $q_{T}, q_{VX}$: The branch selection qubits (2 qubits, outer). 
        \item $f_{\nu}$: The nested boxes failure flag (1 qubit, outer).
        \item Total: 3 qubits.
    \end{enumerate}
    \item[2.] Spatial ($w$) \& Bit-Index ($r, b$):
    \begin{enumerate}
       \item $w$: The spatial dimension (2 qubits, since $n=2$ for 3 states, outer).
       \item $r$: The bit-index (needs $n_p - 1$ qubits for $0$ to $n_p-2$, inner).
       \item $b$: As part of block encoding $T$ (1 qubit, inner).
       \item Total: $n_p + 2$ qubits.
    \end{enumerate}
    \item[3.] Nested Boxes ($|\nu\rangle$):
    \begin{enumerate}
       \item $\nu$: The 3D momentum transfer vector ($3n_p + 3$ qubits, accounting for sign bits, outer).
       \item $\mu$: The box index ($n_p$ qubits in unary, outer).
       \item $m_{\text{nb}}$: The inequality test uniform state ($n_{\mathcal{M}}$ qubits, outer).
       \item Total: 
       We eliminate the inner boxes, unlike in Ref.~\cite{berry2024quantum}, and thus the total is $4n_{p} + n_{\mathcal{M}} + 5$.
    \end{enumerate}
    \item[4.] Shared Index ($\texttt{idx}, f_{\text{nuc-flag}}$):
    \begin{enumerate}
        \item $\texttt{idx}$: The shared index register ($\lceil \log_2 \max(\eta, Z_{\text{tot}}) \rceil$ qubits, outer and inner).
        \item $f_{\text{nuc-flag}}$: The nuclear flag (1 qubit, inner).
        \item Total: $\lceil \log_2 \max(\eta, Z_{\mathrm{tot}}) \rceil + 1$ qubits.
    \end{enumerate}
    \item[5.] Overflow bits for $X$ and $V$ and the flag
     \begin{enumerate}
         \item two ancilla bits for each direction $w$ (6 qubits, outer)
         \item one qubit flagged when the addition overflows in $V$ or $X$ (inner)
     \end{enumerate}
\end{enumerate}
Overall, the outer projection qubits are
\begin{align}
q_{T}, q_{VX}, f_{\nu}, w, \nu, \mu, m, 
\texttt{idx}, \text{overflow-w}.
\end{align}
As a result, the total number of qubits is
\begin{align}
2 + 1 + 2 + 3n_p + 3 + n_p + n_{\mathcal{M}} + n_{\text{idx}} + 6
= 4n_p+ n_{\mathcal{M}}+ n_{\text{idx}} +14.
\end{align}
where $n_{\text{idx}}\equiv\lceil \log_2 \max(\eta, Z_{\text{tot}}) \rceil$.
The cost of a reflection on this number of qubits is 2 Toffolis less than this, though there are 2 extra needed for the phase estimation.
In each step of phase estimation there is one more Toffoli to make this reflection controlled, as well as one more Toffoli for the unary iteration on the control register.
The inner projection qubits are
\begin{align}
&T:  r, b \\
&V:  \texttt{idx}, f_{\text{nuc-flag}}, \text{overflow-flag}\\
&X:  \text{overflow-flag}
\end{align}
As a result the number of qubits to reflect on is
\begin{equation}
    n_p + n_{\text{idx}} +2.
\end{equation}
The reflection cost would be 2 Toffolis less than this, except there is an extra control of the reflection on $\texttt{idx}$, which costs one more Toffoli.
This is to ensure that the inner and outer sum over $j$ are distinguished between the terms of the Hamiltonian.

\subsection{Hardware-Inflated 1-Norms and the Effective $\lambda_{\text{eff}}$}\label{app:inflated_1norm}

To physically execute the SOSSA block encoding on quantum hardware, the ideal analytical 1-norms ($\Lambda_T, \Lambda_V, \Lambda_X$) defined previously must be systematically inflated. This inflation accounts for the finite precision of the state preparation subroutines and the inherent failure probabilities of the quantum arithmetic. We define the true, hardware-targeted 1-norms required to compile the branch selection rotations ($P_{\text{branch}}$) and determine the overall phase estimation complexity.

First, the base 1-norms are inflated by their respective hardware discretizations:
\begin{itemize}
    \item \textbf{Kinetic ($\Lambda'_T$):} The exponentially-weighted unary cascade prepares $2^{n_p-1}$ states instead of the exact physical boundary of $2^{n_p-1}-1$. The inflated 1-norm is $\Lambda'_T = \frac{6\pi^2\eta}{\Omega^{2/3}} 2^{2(n_p-1)}$.
    \item \textbf{Potential ($\Lambda^\kappa_V$) \& Self-Interaction ($\Lambda^\kappa_X$):} The momentum transfer register $|\nu\rangle$ is prepared using discrete nested boxes of precision $\mathcal{M}=2^{n_{\mathcal{M}}}$.
    A scaling factor $\kappa$ (denoted $\alpha$ in Ref.~\cite{su2021fault}) may be adjusted to make the state more accurate.
    The 1-norms with the combined discretisation and scaling factor are denoted $\Lambda^\kappa_V$ and $\Lambda^\kappa_X$.
\end{itemize}
\subsubsection{Discrete Nested Boxes Inflation ($\Lambda^\kappa_V, \Lambda^\kappa_X$):}
Both the Potential ($\tilde{V}_\nu$) and Self-Interaction ($X_{j,\nu}$) operators require preparing the momentum transfer register $|\nu\rangle$ such that the amplitude of each state is strictly proportional to $1/\|\nu\|$. However, the nested boxes state preparation subroutine achieves this via a coherent inequality test against a uniform discrete dummy variable $m \in [0, \mathcal{M}-1]$ (where $\mathcal{M} = 2^{n_{\mathcal{M}}}$). 

Because $m$ is an integer, the resulting preparation probabilities are discretized.
This makes the amplitudes slightly larger than needed, so the error can be reduced by introducing a scaling factor $\kappa\le 1$.
Then the amplitudes may be smaller or larger than required, with the average error being reduced.
The true, hardware-prepared sum over the discrete grid, which we denote $\lambda_\nu^\kappa$, is given by:
\begin{align}
    \lambda_\nu^\kappa = \kappa \sum_{\mu=2}^{n_p+1} \sum_{\nu \in B_\mu} \frac{\lceil \mathcal{M} (2^{\mu-2}/\|\nu\|)^2 \rceil}{\mathcal{M} 2^{2\mu-4}}. \label{eq:lambda_nu_alpha}
\end{align}
In the limit of infinite precision ($\mathcal{M} \to \infty$), the ceiling function vanishes and $\lambda_\nu^\kappa$ perfectly converges to the ideal continuous sum $\lambda_\nu = \sum_{\nu \in G_0} \frac{1}{\|\nu\|^2}$. 

Because the $\tilde{V}_\nu$ and $X_{j,\nu}$ operators share this exact outer $|\nu\rangle$ state preparation, their physical 1-norms on the quantum hardware are strictly proportional to this discrete sum. We define their hardware-inflated 1-norms by scaling the ideal analytical norms ($\Lambda_V$ and $\Lambda_X$) by the ratio of the sum over discretized values to the sum over exact values:
\begin{align}
    \Lambda^\kappa_V = \Lambda_V \left( \frac{\lambda_\nu^\kappa}{\lambda_\nu} \right), \quad \Lambda^\kappa_X = \Lambda_X \left( \frac{\lambda_\nu^\kappa}{\lambda_\nu} \right). \label{eq:inflated_V_X}
\end{align}
By using $\Lambda^\kappa_V$ and $\Lambda^\kappa_X$ in our subsequent compilations, we guarantee that the finite precision of the nested boxes inequality test is  absorbed into the total 1-norm of the block encoding.

\subsubsection{Success probability inflation of 1-norms}
Next, we must account for the asymmetric amplitude loss inherent to the direct-sum architecture. The uniform state preparations for the spatial dimension ($w$), the electrons ($j$), and the nuclei ($\ell$) succeed with high, but imperfect, probabilities: $p_w, p_j,$ and $p_\ell$ (typically $>0.999$). 

In a standard LCU, the target operator is applied once per block encoding. In the SOSSA protocol, the inner reflection $\Pi_{\text{in}}$ occurs \textit{between} the forward and reverse applications of the block-encoding unitary $U_R$ within the walk operator $W$. 
For the kinetic ($T$) and self-interaction ($X$) branches, the electron index $j$ acts as an outer variable. It is prepared once in $P_{\text{out}}$, entirely bypasses $\Pi_{\text{in}}$, and is unprepared in $P_{\text{out}}^\dagger$, penalizing its amplitude exactly once by $p_j$ (and $p_w$ for the $T$ branch). 

Conversely, for the potential branch ($\tilde{V}_\nu$), the $\texttt{idx}$ register acts as an inner variable. 
To determine the success probability, we need to account for the weighting in the preparation of the qubit selecting between the electron and nuclear component.
This qubit will be prepared in the state
\begin{equation}
    \ket{\psi_V} = \frac 1{\sqrt{\eta/p_j+Z_{\text{tot}}/p_\ell}} \left( 
    \sqrt{\eta/p_j}\ket{0} + \sqrt{Z_{\text{tot}}/p_\ell}\ket{1}
    \right) .
\end{equation}
The postselection on the success of the superpositions then gives the subnormalised state
\begin{equation}
    \frac 1{\sqrt{\eta/p_j+Z_{\text{tot}}/p_\ell}} \left( 
    p_j\sqrt{\eta/p_j}\ket{0} + p_\ell\sqrt{Z_{\text{tot}}/p_\ell}\ket{1}
    \right) .
\end{equation}
The controlled operations in the block encoding followed by the projection onto $\ket{\psi_V}$ then gives a success probability
\begin{equation}
    p_{\text{in}, V} = \frac 1{\eta/p_j+Z_{\text{tot}}/p_\ell} \left( 
    \eta + Z_{\text{tot}}
    \right) .
\end{equation}
To survive the intermediate $\Pi_{\text{in}}$ reflection, this inner state must be prepared and unprepared during the forward pass, and then prepared and unprepared \textit{again} during the reverse pass. Consequently, the $V$ branch suffers this combined failure rate twice, yielding a survival amplitude of $(p_{\text{in}, V})^2$. 

To prevent this asymmetric amplitude loss from skewing the simulated Hamiltonian, we statically rebalance the target 1-norms before classical compilation. Crucially, by reweighting the entire $V$ branch uniformly, we perfectly preserve the $(\eta + Z_{\text{tot}})^2$ cross-terms natively generated by the sum-of-squares algebra:
\begin{align}
    \Lambda_{T, \text{target}} = \frac{\Lambda'_T}{p_w p_j}, \quad \Lambda_{X, \text{target}} = \frac{\Lambda^\kappa_X}{p_j}, \quad \Lambda_{V, \text{target}} = \frac{\Lambda^\kappa_V}{(p_{\text{in}, V})^2}.
\end{align}

Lastly, we apply the trash-recycling logic to absorb much of the massive $\sim 75\%$ failure rate of the nested boxes $|\nu\rangle$ preparation ($p_\nu$) when amplitude amplification is not applied.
As detailed above in Appendix \ref{app:be_nu_state_prep}, the kinetic operator does not depend on $\nu$, so $T$ may be block encoded in cases of failed $|\nu\rangle$ preparation.
The hardware-inflated 1-norms described here can be used in the formulae of Appendix \ref{app:be_nu_state_prep} to give the overall $\lambda_{\rm eff}$ for the block encoding.

\section{Jellium block encoding}
Jellium is a simpler system to block encode for the SOSSA walk operator because it no longer involves the nuclear term. We drop the need for a nuclear flag $f_{\text{nuc-flag}}$, QROM for the $\tilde{R}_{\ell}$ coordinates, and phasing of $e^{-i k_{\nu}\cdot R_{\ell}}$. We also simplify the trash overhead recycling of nested boxes now that the LCU 1-norm no longer involves the LCU for the nuclear phasing component.  The differences are as follows:

\begin{enumerate}
    \item \textbf{Elimination of Nuclear Arithmetic:} The nuclear routing flag ($f_{\text{nuc-flag}}$) and the uniform superposition over the nuclei ($Z_{\text{tot}}$) are no longer needed. Consequently, the $\mathcal{O}(n_p n_R)$ shift-and-add multiplier required to accumulate the $e^{-ik_\nu \cdot R_\ell}$ phase is eliminated from the SELECT oracle.
    \item \textbf{Flattened State Preparation:} Because every remaining SOS generator ($T_{j,w}, {V}_\nu, X_{j,\nu}$) targets an electron, the uniform superposition over the $j$-register is required universally. We still need to reflect around the zero state for the $V_\nu$ term for the $s=1~(V)$ branch of the wavefunction.
    That is, there is an inner sum over $j$ for $V_\nu$ but an outer sum for $T_{j,w}$ and $X_{j,\nu}$.
    As with the case with nuclei, rather than making the inverse and forward preparation of the equal superposition for the $j$-register controlled,
    the reflection can be controlled on $\text{Exec}_{V}$. 
    As a result, there is no control of the uniform state preparation needed, and its cost is $3n_{\eta} + 2b_{r} -9$ for each of the prepare-$j$ and unprepare-$j$.
    \item We can redefine our inner LCU 1-norm for the potential term as it no longer involves the nuclear interaction.
    It is then $\lambda_{\tilde{V}_\nu} = \eta$, without the $Z_{\text{tot}}$ term.
\end{enumerate}
The SOS Jellium block encoding normalization is thus
\begin{align}
    \lambda_{\text{SOS}}^{(\text{jellium})} &= \sum_{j=1}^\eta \sum_{w\in \{x,y,z\} }\lambda_{T_{j,w}}^2 + \sum_{\nu\in G_0} C_{\nu}\lambda_{\tilde{V}_{\nu}}^2+\sum_{j=1}^{\eta}\sum_{\nu\in G_0} C_{\nu}\lambda_{\tilde{X}_{j,\nu}}^2 \nn
    &= \frac{3\pi^2\eta}{2\Omega^{2/3}} (N^{1/3}-1)^2 + \eta^{2}\sum_{\nu \in G_{0}}C_{\nu} + \eta\sum_{\nu\in G_{0}}C_{\nu} \nn
    &= \lambda_{T} + \eta\left(\eta + 1\right) \lambda_{\text{out}} .
\end{align}
The difference from the expression for $\lambda_{\text{SOS}}$ in Eq.~\eqref{eq:lambda_sos_asymp} is that there is not the $Z_{\text{tot}}$ term.
Because that is $\mathcal{O}(\eta)$, we obtain the same asymptotic scaling for $\lambda_{\text{SOS}}^{(\text{jellium})}$.
Therefore, the effective 1-norm has the same scaling as in Eq.~\eqref{eq:sossa_lambda_effective}. The lower bound is 
\begin{align}
\beta &= \frac{\eta}{2\pi \Omega^{1/3}} \sum_{\nu \in G_0}\frac{1}{\lVert \nu \rVert^2} 
\end{align}
which scales as  $\mathcal{O}(\eta\Delta^{-1})$.
Therefore, we obtain an asymptotic speedup over the standard LCU method in both the thermodynamic and continuum regimes, similar to the case with nuclei.

\begin{table}[H]
\caption{Costs for \textsc{PREP} operations for the Jellium block encoding} 
\label{tab:prep_costs_jellium}
\begin{ruledtabular}
\begin{tabular}{l c c c c}
\textbf{Subroutine} & $P_{\text{out}}$ & $P_{\text{out}}^{\dagger}$ & \textbf{Total (1 Call)} & \textbf{Calls} $/ W$ \\
\hline
\hline \rule{0pt}{3ex}
Spatial dimensions ($w$) & $13$ & $13$ & $26$ & $1$ \\
Branch prep rotations ($q_T, q_{VX}$) & $2n_T - 6$ & $2n_T - 6$ & $4n_T - 12$ & $1$ \\
Nested Boxes ($|\nu\rangle$, $f_\nu$) &  
$3  n_p^2 + 11 n_p + 2 +  4 n_{\mathcal{M}}  (n_p + 1)$ &  $0$ (Measurement)
& $3n_p^2 + 11n_p +2 + 4n_{\mathcal{M}}(n_p + 1)$ & $1$ \\
Semantic Logic ($\text{Exec}_{T, V, X}$) & $2$ & $0$ (Measurement) & $2$ & $1$ \\
\hline
\textbf{Subroutine} & $P_{\text{in}}$ & $P_{\text{in}}^{\dagger}$ & \textbf{Total (1 Call)} & \textbf{Calls} $/ W$ \\
\hline
Bit-index cascade ($r$) & $n_p - 2$ & $0$ (Measurement) & $n_p - 2$ & $2$ \\
Electron index $(j)$ &  $3n_\eta + 2b_r - 9$ & $3n_\eta + 2b_r - 9$ & $6n_\eta + 4b_r - 18$ & $2$ \\
\end{tabular}
\end{ruledtabular}
\end{table}

\begin{table}[H]
\caption{Comprehensive Toffoli Costs for the walk operator reflections for Jellium including controls for quantum phase estimation.}
\label{tab:reflect_costs_jelllium}
\begin{ruledtabular}
\begin{tabular}{l c c}
\textbf{Reflection Subroutine} & \textbf{Total Toffoli Cost} & \textbf{Calls} $/ W$ \\
\hline
\multicolumn{3}{c}{\textit{Inner Reflection $(2\Pi_{\text{in}} - \mathbb{I})$}} \\
\hline \rule{0pt}{3ex}
\textbf{Kinetic ($E_T$):} Reflect on $r$ and $b$ & $n_p$ & $1$ \\
\textbf{Potential ($E_V$):} Reflect on $\eta$, \text{overflow-flag} & $n_{\eta} + 1$ & $1$ \\
\textbf{Self-Interaction ($E_X$):} 0 & 0 & $1$ \\
\hline \rule{0pt}{3ex}
\textit{Subtotal for Inner Reflection:} & $n_p + n_{\eta}$ & \textit{1 (Sequential Sum)} \\
\hline
\multicolumn{3}{c}{\textit{Global Outer Reflection $(2\Pi_{\text{all}} - \mathbb{I})$}} \\
\hline \rule{0pt}{3ex}
Reflect on all active ancilla registers ($N_{\text{all}}$ qubits) & $4n_p + n_{\mathcal{M}} + n_{\eta} + 14$ & $1$ \\
\end{tabular}
\end{ruledtabular}
\end{table}

\section{Alternative SOS representations}\label{app:alt_sos_reps}

The SOS representation of the electronic structure Hamiltonian (Eq.~\eqref{eq:hamiltonian_terms}) in Eq.~\eqref{eq:sos_decomposition} is not unique.
In this appendix we present two alternative representations.
Recall that to achieve an asymptotic speed-up,  the effective 1-norm $\lambda_{\text{eff}} \leq\sqrt{\lambda_{\text{SOS}} \beta}$ must scale better than the LCU 1-norm in Eq.~\eqref{eqn:lambda_lcu_scaling}, while the block encoding cost remains similar to the LCU approach.
The SOS construction in the main text achieves this by combining the electron-electron term $V$ and the electron-nuclear term $U$ into a single sum of squares.
In Appendix~\ref{app:alt_sos_1} we instead construct separate SOS representations for each term $T$, $U$, and $V$.
In Appendix~\ref{app:alt_sos_2} we consider a decomposition in which the potential term $U$ is combined with the kinetic term $T$. In Appendix~\ref{app:alt_sos_3} we comment briefly on other approaches.

\subsection{Separate SOS decompositions of $T$, $U$, and $V$}
\label{app:alt_sos_1}

The most direct approach is to construct independent SOS representations of the kinetic, interaction, and potential terms.
For the kinetic term $T$, we use the same decomposition as in Eq.~\eqref{eq:t_sos_gen} in the main text:
\begin{align}
    T &= \sum_{j=1}^\eta \sum_{w\in \{x,y,z\} } T_{j,w}^\dagger T_{j,w},\\
    T_{j,w} &= \frac {\sqrt{2}\pi}{\Omega^{1/3}}\sum_{p\in G} |p_w| \ket{p}\bra{p}_j,
\end{align}
which contributes $\Theta(\eta\Delta^{-2})$ to the SOS 1-norm $\lambda_{\rm SOS}$.
For the interaction term, we can use the same decomposition as in the main text for jellium (setting $\zeta_\ell = 0$):
\begin{align}
    V + \beta_V &= \sum_{\nu\in G_0} V_{\nu}^\dagger V_{\nu}+\sum_{j=1}^{\eta}\sum_{\nu\in G_0} X_{j,\nu}^\dagger X_{j,\nu},\\
V_{\nu} &= \sqrt{C_{\nu}}
\left(\sum_{j=1}^\eta \sum_{\substack{p\in G: \\p-\nu \in G}}  \ket{p-\nu} \bra{p}_j\right),\\
X_{j,\nu} &= \sqrt{C_{\nu}}\sum_{\substack{p \in G: \\p-\nu \notin G}}  \ket{p - \nu} \bra{p}_j,\\
\beta_V &= \eta\sum_{\nu \in G_0}C_{\nu} = \mathcal{O}(\eta \Delta^{-1}).
\end{align}
This term contributes $\Theta(\eta^2\Delta^{-1})$ to $\lambda_{\rm SOS}$.
The potential term $U$ is negative, so to express it as a sum of squares we first shift it by its maximal absolute value
\begin{equation}\label{eq:Umax}
    U_{\max} = \max_{\vec x} |U(\vec x)|. 
\end{equation}
We obtain an SOS representation of $U$ as
\begin{align}
    U +\beta_U &= \sum_{j=1}^\eta U_j^\dagger U_j,\\
    U_j &= \sum_{\vec x} \sqrt{U_{\mathrm{max}} + U(\vec x)} \ket{\vec x}\bra{\vec x}_jU_{\mathrm{FFT}}^{\dagger},\\
    \beta_U &= \eta U_{\mathrm{max}} \, .
\end{align}
The real-space potential, $U(x)$, is local in real space, but it is not equal to the bare Coulomb potential due to the finite FFT grid. Instead it is finite and oscillatory near the nuclei. It can be bounded by the integral of a sinc function:
\begin{equation}
    |U(x)| < \frac{2}{\pi}\sum_{\ell}\zeta_{\ell} \int_0^{k_{\mathrm{max}}}dk \frac{\sin(k|x - R_{\ell}|)}{k|x - R_{\ell}|}
\end{equation}
where $k_{\mathrm{max}} = \max_{\nu\in G_0} \|k_{\nu}\|$. To bound the value of $U_{\mathrm{max}}$ we can split $U$ into short range and long range parts as is done typically to evaluate Ewald sums, $U = U^{\mathrm{lr}} + U^{\mathrm{sr}}$,
\begin{align}
    U_{\mathrm{max}} &\leq \max_{x}|U^{\mathrm{lr}}(x)| + \max_{x}|U^{\mathrm{sr}}(x)| \nn
    &\lesssim \max_{x}|U^{\mathrm{lr}}(x)| + \frac{2}{\pi} \zeta_{\ell_{\mathrm{max}}} \max_x\left[
    \int_0^{k_{\mathrm{max}}}dk \frac{\sin(k|x - R_{{\ell}_{\mathrm{max}}}|)}{k|x - R_{{\ell}_{\mathrm{max}}}|} \right] \nn
    &= \mathcal{O}(\Delta^{-1}),
\end{align}
where $\ell_{\mathrm{max}}$ is the index of a nucleus with maximum charge. Here we have assumed a perfectly periodic system, but this assumption can be relaxed. 
The value of $U^{\mathrm{lr}}$ converges exponentially in momentum space so that its maximum value is just a constant.
This feature is what is required by extensivity, since $U_{\mathrm{max}}$ should be constant in the system size if the energy is to be extensive. This means that the scaling of $\beta_U$ is the same as $\beta_V$,
\begin{equation}
    \beta_{U} = \mathcal{O}(\eta \Delta^{-1}).
\end{equation}
This matches the scaling of the momentum-space SOS presented in Sec.~\ref{sec:fq_sos}, but the block encoding will be more complex due to the necessity of block-encoding one of the terms in real space.

\subsection{SOS decomposition of $T+U$}
\label{app:alt_sos_2}

A second natural attempt is motivated by factorization techniques for the Coulomb Hamiltonian originating in early quantum mechanics
\cite{infeld1951factorization}.
For a single electron interacting with a single nucleus of charge $Z$, the Hamiltonian
\begin{equation}
    H = -\frac{1}{2}\nabla^2 - \frac{Z}{r}
\end{equation}
can be factorized as
\begin{equation}
    H + \frac{Z^2}{2} = \frac{1}{2}  \left( \mathbf{\nabla} +  \frac{Z}{r}\mathbf{r} \right)^\dagger \left( \mathbf{\nabla} +  \frac{Z}{r}\mathbf{r} \right).
\end{equation}
The kinetic term arises from the square of the gradient, the constant shift from the square of the second term $\|\frac{Z}{r}\mathbf{r}\|^2 = Z^2$, and the potential from the cross term via $[\mathbf{\nabla}, \frac{\mathbf{r}}{r}] = \frac{2}{r}$.
In this appendix, we develop an analogue of this decomposition for many electrons and nuclei in the plane wave basis -- a joint SOS representation of terms $T+U$ in Eq.~\eqref{eq:hamiltonian_terms}.

\textbf{Single electron and single nucleus.}
We begin with a discrete analogue of the example above, represented in the plane-wave basis by
\begin{align}
    H^{(0)} &= T^{(0)}+U^{(0)},\\
    T^{(0)} &=\frac{1}{2} \sum_{p \in G} \lVert k_p\rVert ^2 \ket{p}\bra{p}, \\
    U^{(0)} &= - \frac{4\pi Z}{\Omega} \sum_{\nu \in G_0} \frac{1}{\lVert k_\nu \rVert^2} \sum_{\substack{p \in G:\\p-\nu \in G}} \ket{p-\nu}\bra{p}.
\end{align}
Motivated by the continuum factorization, we introduce
\begin{align}
    T_w &= \frac{1}{\sqrt{2}} \sum_{p \in G}  k_{p,w} \ket{p}\bra{p},\\
    U_{\nu, w} &= \frac{4\sqrt{2}\pi Z}{\Omega} \frac{k_{\nu, w}}{\lVert k_\nu \rVert^4} \sum_{\substack{p \in G:\\p-\nu \in G}} \ket{p-\nu}\bra{p},\\
    A_{\nu, w} &= \frac{1}{\sqrt{|G_0|}} T_w - \sqrt{|G_0|} U_{\nu, w}.
\end{align}
The operators $A_{\nu, w}$ are defined so that the sum of their squares,
\begin{equation}
    \sum_{w \in \{x,y,z\}} \sum_{\nu \in G_0} A_{\nu, w}^\dagger A_{\nu, w} = \sum_{w \in \{x,y,z\}} T_w^\dagger T_w - \left(\sum_{w \in \{x,y,z\}} \sum_{\nu \in G_0} T_w^\dagger U_{\nu,w} + U_{\nu, w}^\dagger T_w\right) + |G_0|\sum_{w \in \{x,y,z\}} \sum_{\nu \in G_0} U_{\nu,w}^\dagger U_{\nu, w},
\end{equation}
follows the same pattern as in the continuum case.
The kinetic term is recovered from
\begin{equation}
\sum_{w \in \{x,y,z\}} T_w^\dagger T_w = T^{(0)},
\end{equation}
while the cross terms reproduce the Coulomb potential,
\begin{align}
    \sum_{w \in \{x,y,z\}} \sum_{\nu \in G_0} T_w^\dagger U_{\nu,w} + U_{\nu, w}^\dagger T_w &= \frac{4\pi Z}{\Omega}\sum_{\nu \in G_0} \sum_{p \in G}\frac{k_\nu \cdot k_p}{\lVert k_\nu \rVert^4} \sum_{\substack{q \in G:\\q-\nu \in G}}(\ket{p}\braket{p}{q-\nu}\bra{q}+\ket{q}\braket{q-\nu}{p}\bra{p})\\
    &=\frac{4\pi Z}{\Omega} \sum_{\nu \in G_0} \frac{1}{\lVert k_\nu \rVert^2} \sum_{\substack{p \in G:\\p-\nu \in G}} \ket{p-\nu}\bra{p}\\
    &= -U^{(0)}.
\end{align}
The second identity follows from the relation $k_\nu\cdot k_{p-\nu}-k_\nu\cdot k_p=\|k_\nu\|^2$.
Unlike the continuum factorization, the remaining squared term is not proportional to the identity. Instead,
\begin{equation}
     \sum_{w \in \{x,y,z\}} \sum_{\nu \in G_0} U_{\nu,w}^\dagger U_{\nu, w} = \frac{32\pi^2 Z^2}{\Omega^2} \sum_{\nu \in G_0} \frac{\lVert k_\nu \rVert^2}{\lVert k_\nu \rVert^8}\sum_{\substack{p \in G:\\p-\nu \in G}} \ket{p}\bra{p}.
\end{equation}
This diagonal operator is missing contributions from momenta $p$ such that $p-\nu \notin G$.
To complete the factorization, we introduce boundary operators
\begin{equation}
    Y_\nu = \sqrt{|G_0|}\frac{4\sqrt{2}\pi Z}{\Omega} \frac{1}{\lVert k_\nu \rVert^3} \sum_{\substack{p \in G:\\p-\nu \notin G}} \ket{p}\bra{p},
\end{equation}
which fill in the missing diagonal terms.
This yields the SOS decomposition
\begin{equation}
     \sum_{w \in \{x,y,z\}} \sum_{\nu \in G_0} A_{\nu, w}^\dagger A_{\nu, w}  + \sum_{\nu \in G_0} Y_{\nu}^\dagger Y_{\nu} = H+\beta^{(0)}
\end{equation}
with constant shift
\begin{equation}
    \beta^{(0)} = |G_0| \frac{32\pi^2 Z^2}{\Omega^2} \sum_{\nu \in G_0} \frac{1}{\lVert k_\nu \rVert^6}.
\end{equation}
Since $|G_0|=\Theta(\Omega\Delta^{-3})$ and $\frac{1}{\Omega^2}\sum_{\nu \in G_0} \frac{1}{\lVert k_\nu \rVert^6} = \Theta(1)$ as the grid spacing $\Delta\to0$, we have $\beta^{(0)} = \Theta(\Omega \Delta^{-3})$. For fixed density $\Omega$ is scaling linearly with $\eta$ so $\beta^{(0)} = \Theta(\eta \Delta^{-3})$.

\textbf{Multiple nuclei.}
We now generalize the above SOS to $L$ nuclei with charges $\zeta_\ell$ and positions $R_\ell$, represented by the Hamiltonian
\begin{align}
    H^{(1)} &= T^{(1)}+U^{(1)},\\
    T^{(1)} &=\frac{1}{2} \sum_{p \in G} \lVert k_p\rVert ^2 \ket{p}\bra{p}, \\
    U^{(1)} &= - \frac{4\pi}{\Omega} \sum_{\ell=1}^L\sum_{\nu \in G_0} \zeta_\ell \frac{e^{-ik_\nu\cdot R_\ell}}{\lVert k_\nu \rVert^2} \sum_{\substack{p \in G:\\p-\nu \in G}} \ket{p-\nu}\bra{p}.
\end{align}
For each nucleus $\ell$, we define a separate set of operators $A_{\ell, \nu, w}$ to reproduce its contribution to $U^{(1)}$, weighted by positive parameters $w_\ell$:
\begin{align}
    A_{\ell, \nu, w} &= \sqrt{\frac{w_\ell}{|G_0|}}T_{w} - \sqrt{\frac{|G_0|}{w_\ell}} U_{\ell,\nu, w},\\
    U_{\ell,\nu, w} &= \zeta_{\ell}e^{-ik_\nu \cdot R_\ell}\frac{4\sqrt{2}\pi }{\Omega} \frac{k_{\nu, w}}{\lVert k_\nu \rVert^4} \sum_{\substack{p \in G:\\p-\nu \in G}} \ket{p-\nu}\bra{p},\\
    Y_{\ell,\nu} &= \sqrt{\frac{|G_0|}{w_\ell}}\zeta_\ell 
    \frac{4\sqrt{2}\pi }{\Omega} \frac{1}{\lVert k_\nu \rVert^3} \sum_{\substack{p \in G:\\p-\nu \notin G}} \ket{p}\bra{p}.
\end{align}
If the weights $w_\ell$ satisfy $\sum_{\ell=1}^L w_\ell = 1$, by the same arguments as before we have
\begin{align}
     \sum_{\ell=1}^L\sum_{w \in \{x,y,z\}} \sum_{\nu \in G_0} A_{\ell, \nu, w}^\dagger A_{\ell\nu, w}  + \sum_{\ell=1}^L\sum_{\nu \in G_0} Y_{\ell,\nu}^\dagger Y_{\ell,\nu} &= H^{(1)}+\beta^{(1)}, \\
     \beta^{(1)} &= |G_0| \frac{32\pi^2}{\Omega^2} \sum_{\nu \in G_0} \frac{1}{\lVert k_\nu \rVert^6} \sum_{\ell=1}^L\frac{\zeta_\ell^2}{w_\ell}.
\end{align}
Optimizing the weights $w_\ell$ so that the constant shift is minimal gives
\begin{align}
    w_\ell &= \frac{\zeta_\ell}{Z_{\text{tot}}},\\
    \beta^{(1)} &= |G_0|\frac{32\pi^2 Z_{\text{tot}}^2}{\Omega^2} \sum_{\nu \in G_0} \frac{1}{\lVert k_\nu \rVert^6}.
\end{align}
The nuclear positions only enter $H^{(1)}$ through the phases $e^{ik_\nu \cdot R_\ell}$, which cancel when taking the square.
The shift $\beta^{(1)}$ depends on the nuclear configuration only through the total charge, and is identical to the single-nucleus result with $Z \to Z_{\text{tot}}$.

\textbf{Full many-electron Hamiltonian.}
Finally, we construct an SOS representation for the Hamiltonian in Eq.~\eqref{eq:hamiltonian_terms}, with multiple nuclei and electrons.
For single-particle operators, we repeat the construction above independently for each electron:
\begin{align}
    T_j +U_j + |G_0|\frac{32\pi^2 Z_{\text{tot}}^2}{\Omega^2} \sum_{\nu \in G_0} \frac{1}{\lVert k_\nu \rVert^6} = \sum_{\ell=1}^L\sum_{w \in \{x,y,z\}} \sum_{\nu \in G_0} A_{j,\ell,\nu,w}^\dagger A_{j,\ell, \nu, w}+ \sum_{\ell=1}^L\sum_{\nu \in G_0} Y_{j,\ell,\nu}^\dagger Y_{j,\ell,\nu}.
\end{align}
Combining this with the SOS decomposition of the interaction term $V$ from Appendix~\ref{app:alt_sos_1}, we obtain
\begin{equation}
    H + \beta = \sum_{j=1}^{\eta} \sum_{\ell=1}^L\sum_{w \in \{x,y,z\}} \sum_{\nu \in G_0} A_{j,\ell,\nu,w}^\dagger A_{j,\ell,\nu, w}+ \sum_{j=1}^{\eta} \sum_{\ell=1}^L\sum_{\nu \in G_0} Y_{j,\ell,\nu}^\dagger Y_{j,\ell,\nu} + \sum_{\nu \in G_0} V_{\nu}^\dagger V_{\nu}+ \sum_{j=1}^{\eta}\sum_{\nu \in G_0} X_{j,\nu}^\dagger X_{j,\nu}
\end{equation}
with total shift
\begin{equation}
    \beta = |G_0|\eta\frac{32\pi^2 Z_{\text{tot}}^2}{\Omega^2} \sum_{\nu \in G_0} \frac{1}{\lVert k_\nu \rVert^6} + \frac{\eta}{2\pi \Omega^{1/3}} \sum_{\nu \in G_0}\frac{1}{\lVert \nu \rVert^2} = \Theta(\eta^4\Delta^{-3}).
\end{equation}
The dominant contribution to $\beta$ comes from the decomposition of $T+U$ and scales quartically in $\eta$ and cubically in $\Delta^{-1}$.
Consequently, this approach is asymptotically worse than LCU-based methods and other SOS decompositions in this work.

\subsection{Other SOS decompositions}\label{app:alt_sos_3}
Many other SOS representations are possible. For example, approximating the electron-electron interaction as a sum of separable real-space functions naturally leads to an SOS form. The various terms can also be split up further. Taking the electron-nuclear potential as an example, one could imagine treating it partially in momentum space and partially in real space. In general, combining different methods for different terms leads to a large design space of possible SOS decompositions with different properties. The SOS decomposition presented in Sec.~\ref{sec:fq_sos} and carefully analyzed in this paper is a natural choice, particularly within the context of previous LCU approaches. However, it is possible that some other carefully selected SOS could have better constant factors.

\bibliography{bibliography}
\end{document}